\documentclass[aps,prb,twocolumn,floatfix,superscriptaddress,amsmath,amssymb,longbibliography]{revtex4-2}
\usepackage{xcolor,bm,lineno,multirow,times}
\usepackage{physics}
\usepackage{rotating}
\usepackage{verbatim}
\usepackage{graphicx,tabularx}
\usepackage[sort&compress]{natbib}
\usepackage[T1]{fontenc}
\usepackage{soul}
\usepackage{dcolumn}
\usepackage{bm}
\usepackage[version=4]{mhchem}
\usepackage{booktabs}
\usepackage{gensymb}
\usepackage{color}
\usepackage[colorlinks=true,urlcolor=blue,linkcolor=magenta,citecolor=magenta]{hyperref}

\def\ie{{i.e. }}

\allowdisplaybreaks

\begin{document}

\title{Pressure control of magnetic order and excitations in the pyrochlore antiferromagnet MgCr$_{2}$O$_{4}$ }

\author{L. S. Nassar}
\affiliation{School of Physics, Georgia Institute of Technology, Atlanta, Georgia 30332, USA}
\author{H. Lane}
\email{hlane34@gatech.edu}
\affiliation{School of Physics, Georgia Institute of Technology, Atlanta, Georgia 30332, USA}
\affiliation{School of Physics and Astronomy, University of St Andrews, North Haugh, St Andrews, Fife KY16 9SS, United Kingdom}
\author{B. Haberl}
\affiliation{Neutron Scattering Division, Oak Ridge National Lab, Oak Ridge, Tennessee 37831, USA}
\author{M. Graves-Brook}
\affiliation{Neutron Scattering Division, Oak Ridge National Lab, Oak Ridge, Tennessee 37831, USA}
\author{B. Winn}
\affiliation{Neutron Scattering Division, Oak Ridge National Lab, Oak Ridge, Tennessee 37831, USA}
\author{S.M. Koohpayeh}
\affiliation{Institute for Quantum Matter and Department of Physics and Astronomy, The Johns Hopkins University, Baltimore, Maryland 21218, USA}
\affiliation{Department of Materials Science and Engineering, The Johns Hopkins University, Baltimore, Maryland 21218, USA}
\affiliation{Ralph O’Connor Sustainable Energy Institute, The Johns Hopkins University, Baltimore, Maryland 21218, USA}
\author{M. Mourigal}
\affiliation{School of Physics, Georgia Institute of Technology, Atlanta, Georgia 30332, USA}
\date{\today}

\begin{abstract}
MgCr$_{2}$O$_{4}$ is one of the best-known realizations of the pyrochlore-lattice Heisenberg antiferromagnet. The strong antiferromagnetic exchange interactions are perturbed by small further-neighbor exchanges such that this compound may in principle realize a spiral spin liquid (SSL) phase in the zero-temperature limit. However, a spin Jahn-Teller transition below $T_{\rm N} \approx 13$~K yields a complicated long-range magnetic order with multiple coexisting propagation vectors. We present neutron scattering and thermo-magnetic measurements of MgCr$_{2}$O$_{4}$ samples under applied hydrostatic pressure up to $P=1.7$~GPa demonstrating the existence of multiple close-lying nearly degenerate magnetic ground states. We show that the application of hydrostatic pressure increases the ordering temperature by around 0.8~K per GPa and increases the bandwidth of the magnetic excitations by around 0.5~meV per GPa. We also evidence a strong tendency for the preferential occupation of a subset of magnetic domains under pressure. In particular, we show that the $\boldsymbol{k}=(0,0,1)$ magnetic phase, which is almost negligible at ambient pressure, dramatically increases in spectral weight under pressure. This  modifies the spectrum of magnetic excitations, which we interpret unambiguously as spin waves from multiple magnetic domains. Moreover, we report that the application of pressure reveals a feature in the magnetic susceptibility above the magnetostructural transition. We interpret this as the onset of a short-range ordered phase associated with $\boldsymbol{k}=(0,0,1)$, previously not observed in magnetometry measurements.
\end{abstract}

\maketitle

\section{Introduction}
\label{Sect:Introduction}

Quantum and geometrically frustrated magnetic insulators are ideal platforms to study and understand the interplay between crystallography, statistical mechanics, and many-body quantum theory. The staggering number of compounds and distinct chemical families  available for detailed experimental studies has uncovered a wealth of magnetic ground states and excitations at low temperatures, often stabilized and controlled by small relative energy scales in the material's magnetic Hamiltonian. In spite of this enormous complexity, several key theoretical models, and their approximate material realizations, are attracting continuous attention. One such system is the pyrochlore Heisenberg antiferromagnet (PHA), a paradigm of high magnetic frustration~\cite{Anderson56:102,Lacroix:book}. On the theoretical front, even in the limit where magnetic moments are treated as large spin $S\gg1$ (i.e. classical dipoles), both the system's ground-state and excitations are unusual. The former is governed by an extensive ground-state degeneracy in the $T\rightarrow0$ limit, with spin correlations decaying algebraically in real space such that sharp ``pinch-point'' features appear in reciprocal space~\cite{Reimers92:45,Moessner98:80,Henley10:1}. This is an example of a classical spin-liquid~\cite{Yan23:preprint1}, where the under-constrained interaction rule between spins maps onto an emergent form of magnetic matter, the Coulomb Phase. The spin dynamics in that phase comprises diffusive and precessional dynamics~\cite{Conlon09:102,Zhang2019}. These unusual magnetic properties motivate experimental investigations of materials realizing the PHA. Well-known approximate realizations include the rare-earth fluorides NaA,$^\prime$B$_2$F$_7$ fluorides~\cite{Krizan14:89,Plumb19:15}, the spin-ices~\cite{Reimers92:45,Moessner98:80,Henley10:1} (although magnetic anisotropy dominates in these systems), and spinels such as ZnCr$_2$O$_4$~\cite{Lee02:418}, and MgCr$_2$O$_4$~\cite{Tomiyasu08:101,Dutton11:83,Tomiyasu13:110,Bai19:122}. The latter compound is the subject of this work. 

MgCr$_{2}$O$_{4}$ crystallizes in the $Fd\overline{3}m$ space group with magnetic Cr$^{3+}$ ions on the B sites of the spinel structure which form a pyrochlore lattice. The Cr$^{3+}$ $(S=3/2,l=0)$ ions occupy edge-sharing octahedra of O$^{2-}$ ions which provide a $\approx 90^{\circ}$ Cr-O-Cr superexchange pathway while the short bond-distance in spinel oxides typically allows for a dominant direct $d$-$d$ exchange~\cite{Efthimiopoulos18:97}. As a result, the largest magnetic interaction is the antiferromagnetic nearest-neighbor exchange along the bonds that form the pyrochlore lattice. The strong antiferromagnetic interactions  $|\Theta_{\rm W}| \approx 400$~K in MgCr$_{2}$O$_{4}$ lead to a highly frustrated cooperative paramagnetic regime, that is only relieved at  $T_{\rm N} \approx 13$~K when the compound undergoes a spin Jahn-Teller transition that results in a small tetragonal lattice distortion~\cite{Tchernyshyov02:88,Ortega-San-Martin08:20}. The magnetostructural transition results in complex magnetic ordering that coalesces in the coexistence of three propagation vectors in the ultimate low-temperature N{\'e}el-ordered phase. The magnetic ordering occurs through two distinct transitions from the paramagnetic state~\cite{Shaked70:1}. First, a transition into the ``H-phase'' at around $T\approx16$~K that is partially ordered with  $\boldsymbol{k}_{\rm H}=(0,0,1)$. Then, a transition into the ``L-phase'' is initiated by the magnetostructural transition at  $T\approx13$~K and results in the final long-range ordered structure. This transition has two ordering wave vectors associated with it, $\boldsymbol{k}_{\rm L,1}=(0.5,0.5,0)$ and $\boldsymbol{k}_{\rm L,2}=(0.5,0,1)$, prompting questions whether the low-temperature structure is multi-$\boldsymbol{k}$ or multi-domain. A recent report on extensive diffraction work heavily favors a multi-domain solution~\cite{Gao18:97}. 

The magnetic excitations of MgCr$_{2}$O$_{4}$ above and below the magnetostructural transition have been investigated in detail~\cite{Tomiyasu08:101,Tomiyasu13:110,Gao18:97,Bai19:122}. In the ordered phase, the excitations have been characterized by some authors as resonance-like modes at distinct energies~\cite{Tomiyasu08:101,Tomiyasu13:110}, resembling that of ZnCr$_2$O$_4$~\cite{Lee02:418}, and with a momentum-space structure-factor suggesting the formation of localized spin clusters on hexagons and tetrahedral-pairs of the pyrochlore lattice. A recent report, Ref.~\cite{Bai19:122}, by some of us studied the excitations of MgCr$_2$O$_4$ in the paramagnetic regime. It provided definitive evidence for the presence of subleading further-neighbor Heisenberg exchange interactions in the compound, consistent with the absence of pinch-point scattering in the structure factor~\cite{Conlon10:81}. The former study concluded that the localized spin clusters  are not necessary to explain the dynamics of MgCr$_2$O$_4$ in the paramagnetic phase and mimic the spin correlations produced by further-neighbor Heisenberg exchange interactions. It also hinted that the Hamiltonian of cubic MgCr$_{2}$O$_{4}$ (before the structural distortion) lies close to a spiral spin-liquid ground state~\cite{Gao17:13,Bai19:122}, a classical spin-liquid distinct from the Couloumb phase. This proximity may explain why several magnetic ground states compete and a magnetostructural transition is necessary to relieve frustration at low temperatures. However, due to the complexity of the ultimate magnetic structure, a complete model of the static and dynamic properties of the ordered compound has yet to be determined. 

In this paper, we employ hydrostatic pressures up to $P=1.7$~GPa on polycrystalline and single-crystalline samples of MgCr$_2$O$_4$ to further investigate the nature of the magnetic order and dynamics in the low-temperature phase of this compound. We hypothesize that pressure can tune the relative importance between competing nearest-neighbor exchanges, further-neighbor interactions, and magneto-elastic coupling and conduct comprehensive thermomagnetic and neutron scattering studies to better understand the microscopic origin of the complex magnetic phenomena in this compound. Very much like a recent diffraction report~\cite{Gao18:97}, our work favors a multi-domain magnetically ordered state and demonstrates unambiguously, and in a model-free fashion, that resonance-like excitations in the ordered phase are overlapping spin-waves originating from multiple magnetic domains.

This paper is organized as follows. Sec.~\ref{Sect:Methods} contains experimental details of our joint thermo-magnetic and neutron characterization of polycrystalline and single crystalline samples of MgCr$_2$O$_4$ in applied hydrostatic pressures up to 1.7~GPa. Sec.~\ref{Sect:Results} presents our results and demonstrates pressure control of the magnetic order and excitations in MgCr$_2$O$_4$. Sec.~\ref{Sect:Discussion} discusses the implications of our results on the nature of the low-temperature ground-state in MgCr$_2$O$_4$ as well as mechanistic implications of stabilizing a multidomain magnetic order. This section also provides a short conclusion. A short discussion on systematic effects in our magnetometry experiments under pressure is provided in App.~\ref{sec:appendix-hydrostatic}. More details along with replication results on different samples are provided in supplemental information.

\section{Methods}
\label{Sect:Methods}

\subsection{Synthesis and X-ray Diffraction} 

Polycrystalline samples of MgCr$_{2}$O$_{4}$ were prepared through solid-state synthesis in accordance with the methods of Ref.~\onlinecite{Dutton11:83}. Stoichiometric ratios of magnesium oxide (99.99\%)  and chromium (III) oxide (99.997\%) were thoroughly ground and pressed into a pellet prior to heating in an alumina crucible at 800~$^{\circ}$C for 12 hours followed by firing at 1200~$^{\circ}$C for 24 hours in air. This process was repeated three times before phase purity was achieved as determined from powder x-ray diffraction (PXRD) using a Rigaku Smartlab SE diffractometer operated with K$\alpha$-radiation ($\lambda\!=\!1.5406$~\AA). Samples from two distinct batches, labelled $\#1$ and $\#2$ in the following,  were used in this work [See. Supplementary Figure~\ref{fig:si-pxrd} for PXRD patterns yielding {$a=8.332(1)$~\AA} for the cubic Fd$\bar{3}$m space-group].

{Separate from these samples, our experiments utilized single crystal samples grown at Johns Hopkins University using the optical floating zone technique, as detailed in Ref.~\onlinecite{Koohpayeh13:385}.} The crystal used in this work was chosen and dismounted from the multi-sample assembly used in Ref.~\onlinecite{Bai19:122}. The sample was aligned with a Multiwire Laue backscattering x-ray machine and cut to the required dimensions using a diamond saw. The sample orientation was chosen to maximize the sample volume, given the constraints of the crystal growth axis and the pressure cell geometry.

\subsection{Magnetization Measurements}

Temperature-dependent magnetization $M(T)$ curves of our polycrystalline samples were collected using a Quantum Design MPMS3 magnetometer operated in DC SQUID mode between $T\!=\!1.8$ and $300$~K. High-pressure measurements used a BeCu pressure cell manufactured by HMD (Japan) and polytetrafluoroethylene (PTFE) sample holders. Ambient pressure measurements used the uncompressed pressure cell or a standard brass rod with plastic capsules. Magnetic fields of $\mu_0H=0.01$~T and $0.5$~T were applied, and magnetic susceptibility was calculated as $\chi(T)=M(T)/H$ after background subtraction from the raw magnetization curves. Measurements were conducted under field-cooled (FC, sample cooling down) and zero-field-cooled (ZFC, sample warming up) conditions. Pressure was applied by symmetrically rotating the two threaded pistons of the cell leading to compressions up to $\Delta \ell=2.2$~mm. From the uncompressed state, $\Delta \ell=0$~mm, full pressurization was typically achieved in three steps with intermediate measurement runs after each compression.

Applied pressure $P$ was determined in-situ using the superconducting transition of Pb powder (99.9\%, $T_{\rm c}\approx 6.7$~K in our conditions), added to our samples [See Supplementary Information for further details]. Pb was deemed a good low-temperature manometer given its well-known $T_{c}(P,H)$ dependence~\cite{Suresh07:75} and the proximity to the sample magnetic ordering transition at $T_{\rm N}\approx 13$~K. Measurements taken at $\mu_0H=0.01$~T allowed the diamagnetic response of Pb below $T_c$ to dominate over the paramagnetic signal of our sample. Measurements characterizing our sample were taken above $\mu_0H>0.1$~T thus quenching the transition in Pb and making its contribution negligible. The change in transition temperature between compressions, $\Delta T_c (\Delta \ell)$, was measured from the onset of strong diamagnetism to minimize systematic uncertainty between runs [See Suppl. Fig.~\ref{fig:si-pbtc} for several examples of our in-situ Pb superconducting transition measurements].

{In order to achieve pressures up to $P=1.3$~GPa, polycrystalline MgCr$_{2}$O$_{4}$ was gently packed into the PTFE sample holder until approximately 70-80\% of the available volume was filled. Pb powder was added on top, and Daphne 7373 oil was employed as a pressure-transmitting medium (PTM). The oil and powder assembly was gently emulsified and care was taken to minimize any trapped air during this process. Such a high volume fraction of sample with respect to PTM requires investigation in terms of hydrostaticity and a benchmark of these ``pseudo''-hydrostatic conditions is discussed in  Appendix~\ref{sec:appendix-hydrostatic}. With this understanding and to distinguish from fully non-hydrostatic conditions without a PTM, the pressure will be referred to as hydrostatic for the remainder of the paper.}

Background data were collected using the same pressure cell loaded with an empty PTFE sample holder. The same sequence and parameters of primary data collection were also employed. Raw curves of SQUID voltage versus pressure-cell position were subtracted point-to-point between the sample and background measurements. The resulting curves were analyzed using the SQUIDLab package~\cite{Coak20:91} to extract magnetization values in absolute units using the known instrument-specific calibration factor. In order to minimize systematic uncertainties and artifacts in our data, care was taken to confirm results by repeating experiments. For a comparison of results across replicate experiments, interested readers are directed to Suppl. Fig.~\ref{fig:si-replicates}. 

\begin{figure}[!]
 \centering
    \includegraphics[width=0.99\columnwidth,trim=0cm 0cm 0cm 0cm,clip=true]{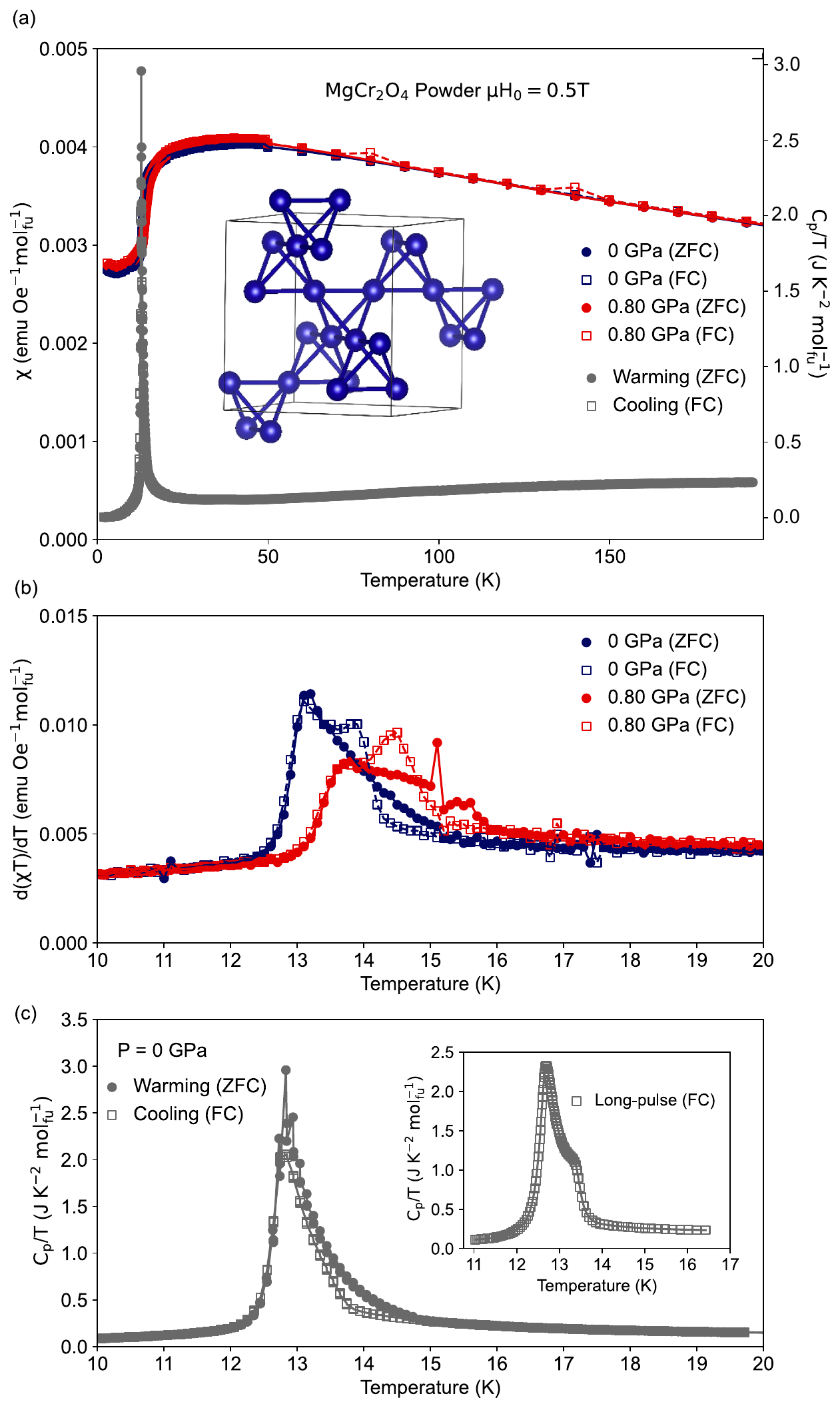}
    \caption{(a) Temperature dependence of the magnetic susceptibility of MgCr$_2$O$_4$ under ambient (blue) and pressurized (red) conditions, as well as field-cooled (FC, cooling down, open symbols) and zero-field cooled (ZFC, warming up, full symbols). Corresponding temperature dependence of the specific heat (grey symbols). Inset: crystal structure of MgCr$_{2}$O$_{4}$ in the high temperature cubic phase, showing only the Cr$^{3+}$ ions and nearest neighbor bonds. (b) Representative plot of the derivative  $\partial [\chi\!\cdot\!T]/\partial T\vert_H$ in the vicinity of $T=T_{\rm N}$ showing the magnetic ordering transition under non-pressurized (blue points) and pressurized (red points) conditions corresponding to $P=0.8 {\rm GPa}$. (c) Temperature dependence of the specific heat measured quasi-adiabatically (short pulse) under ambient pressure and a field of $\mu_0 H =0.5$~T while scanning the same transition warming and cooling. The inset shows the results of a long-pulse measurement over the same temperature range. All data presented here utilized sample $\#1$. See Suppl. Fig.~\ref{fig:si-replicates} for additional independent measurement runs on this sample and others.}
    \label{fig:1}
\end{figure}

\subsection{Heat Capacity Measurements}

{Ambient pressure heat-capacity measurements using a Quantum Design PPMS calorimeter were performed on polycrystalline samples, lightly pressed into solid thin pellets.} Measurements were taken using the standard relaxation method with quasi-adiabatic short ($\leq3\%$ increase) and long ($>30\%$ increase) temperature pulses. The long-pulse results were analyzed using the LongPulseHC package~\cite{Scheie18:193}. The data were converted in specific heat using the known mass ($\pm 5$\%) of the sample.

\subsection{Inelastic Neutron Scattering}
Inelastic neutron scattering measurements were performed on the hybrid time-of-flight spectrometer HYSPEC at the Spallation Neutron Source (Oak Ridge Nationa Laboratory, U.S.). A rod-like single crystal of MgCr$_{2}$O$_{4}$ of mass {$m=0.9$~g}, 9~mm tall and 3.2 mm diamter, cut with horizontal top and bottom facets, was mounted in a clamp cell with fluorinert as a PTM. The pressure cell comprises an aluminum alloy body surrounding a copper-beryllium inner sleeve~\cite{Podlesnyak2018} containing a PTFE sample tube. Tungsten carbide pistons are used to supply pressure to the PTM. The pressure cell was pressurized using a hydraulic press with a three-ton load corresponding to a nominal pressure of $P=1.7$~GPa. Load was transferred from the press to the cell through clamping the top nut and special care was taken to ensure the load was successfully transferred. {Several recent experiments using these cells indicate that pressure loss due to friction is less than 10\% under these conditions. A preliminary experiment using a similar shaped crystal of MgCr$_2$O$_4$ and the same set-up was performed in order to ensure the crystal survived the pressurization event intact.} The pressure cell was shielded with Cd and attached to a bottom-loading close-cycle refrigerator. The sample was aligned with the $c^{*}$ axis of the cubic structure perpendicular to the scattering plane. A slight out-of-plane tilt of $4.8^{\circ}$ was accounted for during the data reduction process. 

The sample was measured at both base temperature ($T = 4.5$~K) and above the sample's magnetic ordering temperature ($T = 20$~K) for both zero compression (nominal pressure $P=0$~GPa) and compressed (nominal pressure $P=1.7$~GPa). HYSPEC was operated with incident energies of $E_{i}$ = 12~meV and $E_{i}$ = 25~meV and a chopper speed of 180 Hz. The HYSPEC detector bank was positioned to cover low scattering angles $2\theta = -3^\circ$ to $-63^\circ$, and the pressure cell was rotated over $180^\circ$ in steps of $1^\circ$. An empty pressure cell was measured at sample rotations of $-90^{\circ}$, $0^{\circ}$ and $90^{\circ}$ and averaged to account for any inhomogeneities in the Cd shielding. The resulting background was then subtracted from the sample data. Data were reduced and analyzed using the Mantid package~\cite{Savici:22:55,Arnold14:764}. The resulting neutron scattering intensity $\tilde{I}({\bf Q},E)=k_i/k_f \left[{\rm d}^2\sigma/{\rm d}\Omega{\rm d}E_f\right]$ as a function of energy transfer $E$ and momentum transfer ${\bf Q} = H \boldsymbol{a}^\ast + K \boldsymbol{b}^\ast + L \boldsymbol{c}^\ast$ is plotted in the cubic setting associated with the high temperature $Fd\overline{3}m$ space group. The vertical focusing on HYSPEC introduces a sizable out-of-plane angular integration with a lower-bound of $\pm 2.4^\circ$ given by the geometry of the guide and focusing array. {For $E_{i}$ = 12~meV, our set-up yielded estimated full-width at half-maximum (FWHM) energy resolutions of $\Gamma (4~{\rm meV})=0.34$~meV and $\Gamma(8~{\rm meV})=0.18$~meV.}


\begin{figure}
 \centering
    \includegraphics[width=0.96\columnwidth,trim=0cm 0cm 0cm 0cm,clip=true]{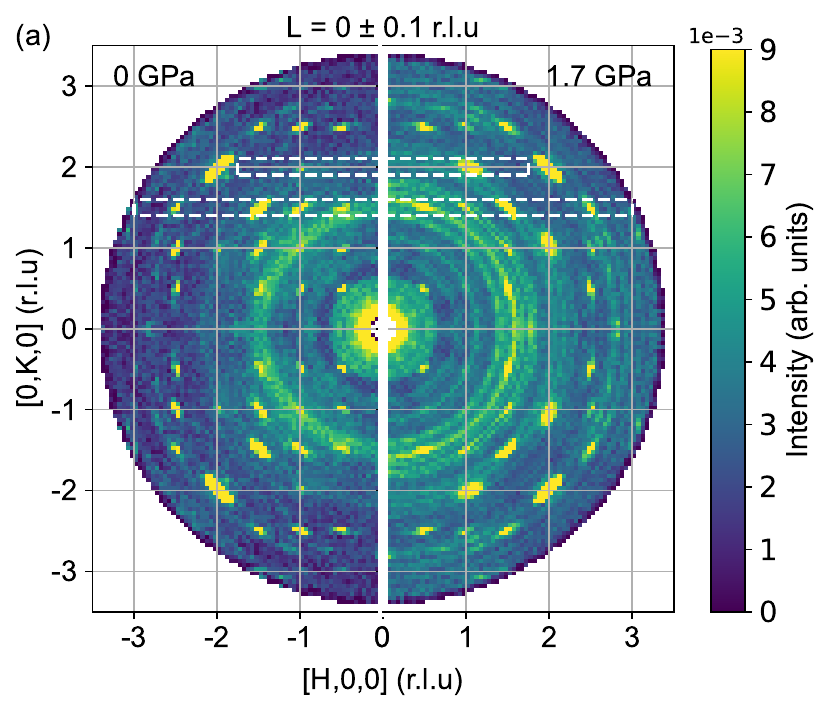}
    \includegraphics[width=0.96\columnwidth,trim=0cm 0cm 0cm 0cm,clip=true]{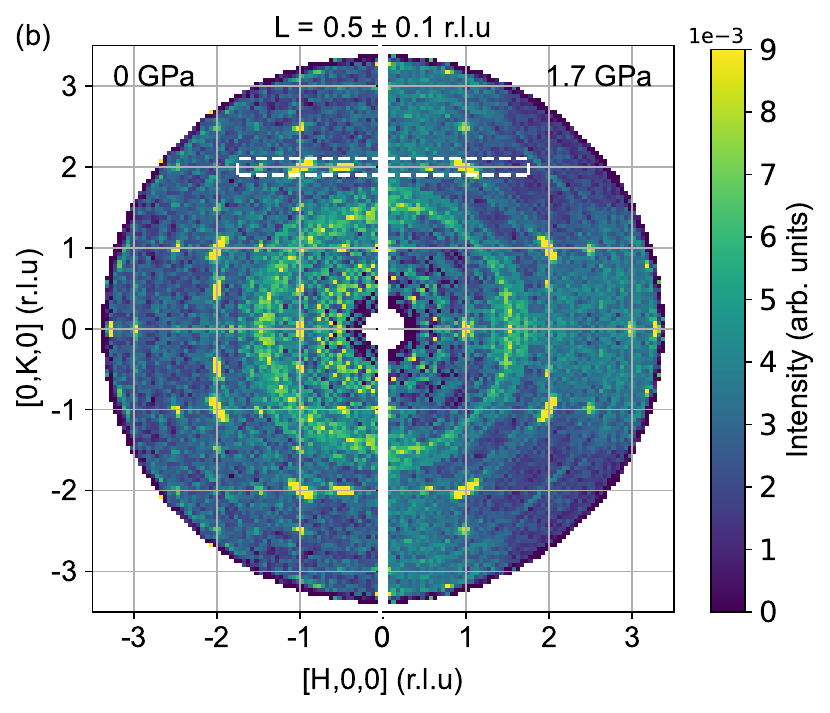}
    \caption{Elastic neutron scattering intensity $\tilde{I}({\bf Q},\bar{E}=0)$ measured with $E_i = 12$~meV at $T\!=\!4.5$~K projected in the $(H,K,\bar{L})$ plane after integrating over $\bar{E}=0\pm 1.0$~meV and (a) $\bar{L} = 0 \pm 0.1$~r.l.u and (b) $\bar{L} = 0.5 \pm 0.1$~r.l.u.  {Such $L$ integration corresponds to an angular integration of $\approx\!2^\circ$ that matches the intrinsic out-of-plane integration on HYSPEC}. Ambient pressure $P=0$~GPa is on the left while $P=1.7$~GPa is on the right. Although the background is subtracted, the intensity is difficult to normalize to absolute units, and the results are left in arbitrary units. The  data were symmetrized in the $m\overline{3}m$ Laue class of the crystal structure before the tetragonal ($c\neq a\!=\!b$) magneto-structural transition at $T_{\rm N}\approx 13$~K, which lowers the crystal symmetry. Given the relaxed vertical resolution on HYSPEC we apply only the symmetry operations in the scattering plane. The dashed white rectangles refer to the one-dimensional cuts presented in Fig.\ref{fig:3}}
    \label{fig:2}
\end{figure}

\section{Results}
\label{Sect:Results}

\subsection{Thermomagnetism of the magnetostructural transition}
\label{Sect:Res-thermo}

To understand the effects of hydrostatic GPa-scale pressure on the magnetic properties of MgCr$_2$O$_4$, we first turn to thermomagnetic results collected on polycrystalline samples inside the pressure-cell. In absence of compression (\ie $P\approx0$), the temperature-dependent susceptibility $\chi(T)$ displays a characteristic Curie-Weiss (CW) behavior $\chi(T) = C/(T-\Theta_{\rm W}) + \chi_0$ above $T^\ast\approx 50$~K [Figure \ref{fig:1}(a)], where $\chi_0$ is the temperature-independent contribution from the sample and the PTM. Given that $|\Theta_{\rm W}|\gtrsim400$~K for MgCr$_2$O$_4$, we did not attempt to extract values for $C$ or $\Theta_{\rm W}$ but we note the good agreement between our data and a previous report from stoichiometric samples~\cite{Dutton11:83}. 

\begin{figure}
\centering
    \includegraphics[width=1.00\columnwidth,trim=0cm 0cm 0cm 0cm,clip=true]{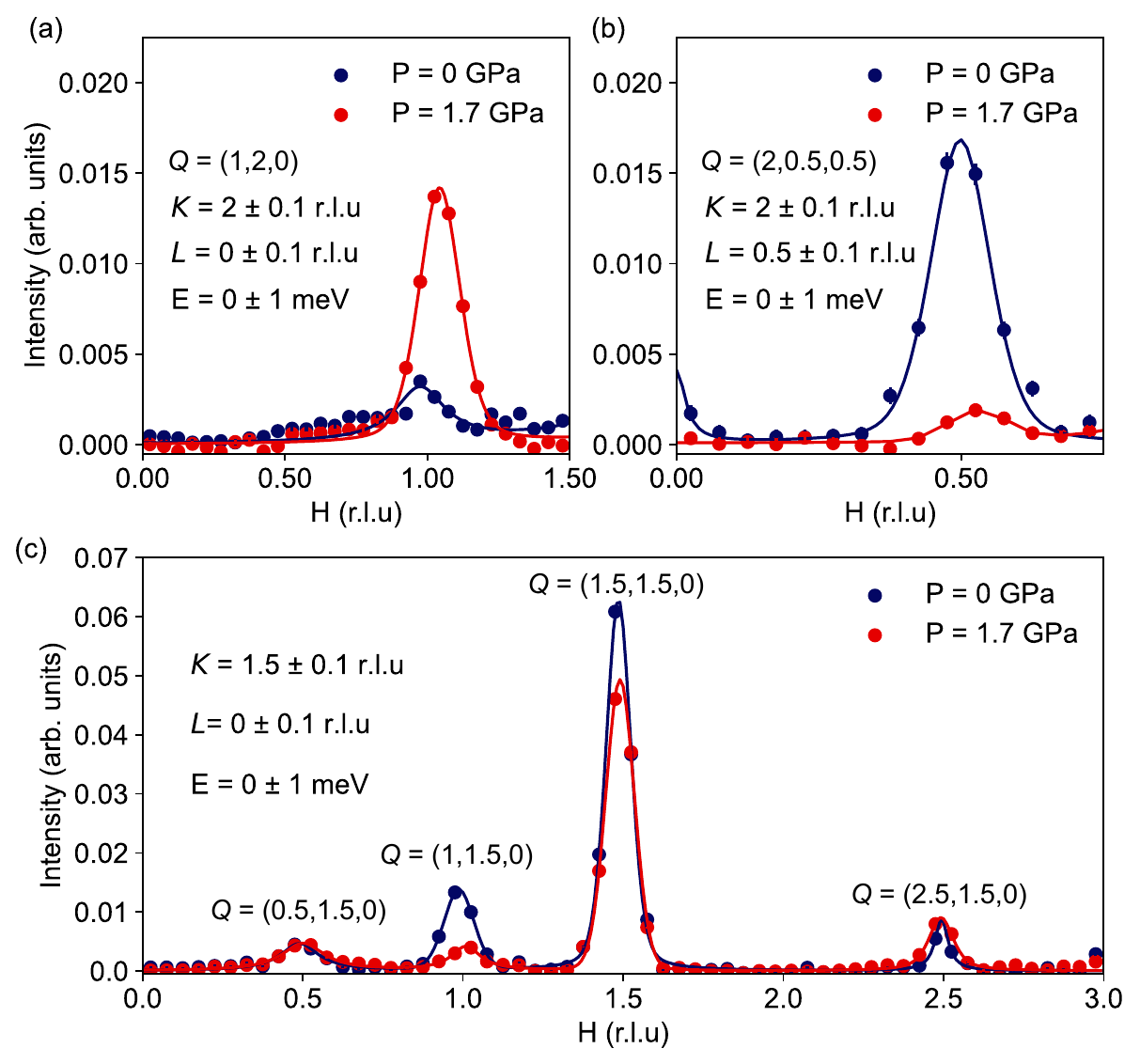}
    \caption{One-dimensional cuts through the $T=4.5~{\rm K}$ elastic data of Fig.~\ref{fig:2} highlighting the redistribution of Bragg peaks between unpressurized (blue circles) and $P=1.7~{\rm GPa}$ (red circles) conditions. Integration ranges are indicated on the plots which correspond to cut directions (a) $\bar{\bf Q}=(H,2,0)$, (b) $\bar{\bf Q}=(H,2,0.5)$,  and (c)  $\bar{\bf Q}=(H,1.5,0)$. Momentum integration ranges are indicated on the plots. The integrated Bragg peak intensity was obtained by fitting to a collection of pseudo-Voigt functions with a linear/quadratic background which was then subtracted from the data.}
    \label{fig:3}
\end{figure}

As is typical for a highly frustrated magnet in the cooperative paramagnetic regime, strong deviations from the CW law are only observed at relatively low temperatures, here for $\tilde{T}\approx 50$~K ($\tilde{T}/|\Theta_W|\approx10$), where  $\chi(T)$ flattens out before experiencing a sharp downturn between $T\!=\!20$ and $10$~K. We associate the sharpest part of this downturn with the magneto-elastic and magnetic ordering phase transition(s) of the sample around $T_{\rm N}\approx13$~K. Correspondingly, the specific heat $C_p(T)$ displays a large $\lambda$-shape peak in agreement with the thermodynamic nature of the transition. A pressure of $P\!=\!0.8$~GPa [Fig.~\ref{fig:1}(a), red points], does not yield significant changes in the overall shape and behavior of the $\chi(T)$ curve. This suggests that the application of GPa-scale hydrostatic pressure does not significantly alter the value of $|\Theta_W|$ or the nature of the cooperative paramagnetic state in MgCr$_2$O$_4$.

To further investigate if there is any effect of hydrostatic pressure on the sample's magnetism, we turn to the behavior of the $\chi(T)$ in the vicinity to $T\approx T_{\rm N}$. To enhance the signature of the magnetic phase transition, we plot $\partial [\chi\!\cdot\!T]/\partial T\vert_H$ [Fig.~\ref{fig:1}(b)] as successfully employed in Ref.~\onlinecite{Dutton11:83} to precisely assign the transition temperature of ACr$_2$O$_4$ spinels from susceptibility measurements. In absence of compression, a pronounced and dominant peak is observed in the ZFC data at $T_{\rm N}(P\approx 0)=13.1(1)~{\rm K}$ with a tail extending up to around $T\approx15~{\rm K}$. The main peak has a slightly higher temperature than the peak observed in specific heat [Fig.~\ref{fig:1}(c)], consistent with Ref.~\onlinecite{Dutton11:83}. The asymmetric peak and overall width, however, is not expected under ambient pressure. {Such lineshape effect typically correlates with a distribution of magnetic transition temperatures in the sample resulting from pressure, temperature and/or compositional gradients. We investigate this effect in detail in Appendix~\ref{sec:appendix-hydrostatic} and conclude it is a systematic effect from the PTM.}

At maximum compression, corresponding to $P=0.8~{\rm GPa}$ for the experiment shown in Fig.~\ref{fig:1}(b), the ZFC peak in $\partial [\chi\!\cdot\!T]/\partial T\vert_H$ shifts by more than $0.5~{\rm K}$ with $T_{\rm N}(P\!=\!0.8~{\rm GPa})$ $=13.7(1)~{\rm K}$ while decreasing in amplitude and broadening substantially with the appearance of two additional features: a weak broad peak at  $T^\prime=15.5$~K, and a sharp single-point peak at $T^{\prime\prime}=15.1$~K. We comment on these features in turn. First, the overall shift, which corresponds to $\partial T_{\rm N} / \partial P \approx 0.75 {\rm K/GPa}$, is continuous as a function of applied pressure [See Suppl. Fig.~\ref{fig:si-replicates} for intermediate pressures for this and replication runs reaching up to $P=1.3~{\rm GPa}$]. We postulate the shift originates from minute changes in further-neighbor exchange interactions and/or magnetoelastic coupling energy, ultimately stabilizing the ordered state at a slightly higher temperature. Second, we associate the peak at $T^\prime$ with a signature of the first transition in MgCr$_2$O$_4$, a phenomenon enhanced under pressure. Third, the sharp discontinuity at $T^{\prime\prime}$ is intriguing as it is systematically present in our measurements in which it gradually appears under pressure [See Suppl. Fig.~\ref{fig:si-replicates}], and always corresponds to a single measurement point, even when scanned with a temperature resolution of $\Delta T=10~{\rm mK}$ [See Suppl. Fig.~\ref{fig:si-discontinuity}]. Despite our efforts, the intrinsic or extrinsic origin of this peak remains unknown.

Comparing FC and ZFC measurements [Fig.~\ref{fig:1}(b)] reveals a significant difference and hysteresis between protocols: the single asymmetric ZFC peak discussed above transforms into a two-peak structure under FC conditions. The second peak at $T=T^\ast$, separated from the main peak at $T=T_{\rm N}$ by $[T^\ast-T_{\rm N}](0~{\rm GPa}) = 0.9(1)~{\rm K}$, shifts and grows in amplitude under pressure with $[T^\ast-T_{\rm N}](0.8~{\rm GPa}) = 0.7(1)~{\rm K}$. The second peak in FC measurements is consistently present in our replication datasets [See Suppl. Fig.~\ref{fig:si-replicates}]. To further investigate this feature from the magnetometry data, we turn to heat capacity measurements at an applied field of $\mu_0H=0.5$~T and under ambient pressure. A subtle hysteresis between cooling (FC) and warming (ZFC) conditions is observed in the specific heat [Fig.~\ref{fig:1}(c)]. This may be expected given that the magnetic ordering in MgCr$_2$O$_4$ is coupled to a first-order structural phase transition, however the second peak is not observed under these conditions. Since quasi-adiabatic heat capacity measurements are not sensitive to the latent heat of first-order phase transitions~\cite{Scheie18:193,Lashley03:43}, we turn to long-pulse measurements [Fig.~\ref{fig:1}(c) inset]. This reveals the second peak upon cooling, which was previously obscured by thermal averaging in the quasi adiabatic protocol. In order to determine the thermal- or field-dependent origin of the second peak, we turn to heat capacity measurements in zero field and pressure [Suppl. Fig.~\ref{fig:si-hc}] which yield the same results, indicating that the second peak in FC $\partial [\chi\!\cdot\!T]/\partial T\vert_H$ results from cooling the sample across $T_{\rm N}$. Based on this evidence, we postulate that the two-peak structure of the transition of MgCr$_2$O$_4$ is likely present at ambient pressure, but its signature is enhanced by the application of hydrostatic pressure upon cooling.

In short, our careful thermomagnetic measurements reveal that the magnetoelastic transition temperature(s) in MgCr$_2$O$_4$ can be shifted by almost $0.8~{\rm K}$ per GPa while the overall susceptibility behavior remains unchanged. Perhaps more importantly to understand the nature of that transition, we observe subtle lineshape and double-peak structure effects extending by almost $2~{\rm K}$ above the main transition at $T_{\rm N}$. These effects are impacted and enhanced by applied pressure but can be traced back to ambient pressure measurements. Given the known multi-structure and multi-domain propensity of  MgCr$_2$O$_4$, we hypothesize these effects originate from magnetic domain control and redistribution. We turn to elastic and inelastic neutron scattering measurements to gain a direct microscopic view of these effects.

\subsection{Neutron Scattering}
\label{Sect:Res-neutrons}

\subsubsection{Elastic Scattering}

We now turn to results from elastic neutron scattering collected at $T=4.5$~K, well below $T_{\rm N}$, and compare ambient pressure with results for $P=1.7$~GPa. Two-dimensional slices through the  $m\overline{3}m$-symmetrized elastic data are presented in Fig.~\ref{fig:2}, displaying Bragg peaks associated with the structural and magnetic long-range order. Structural Bragg peaks corresponding to the $\{220\}$-family of the cubic structure are visible for ${\bf Q}  = \boldsymbol{\tau}_{\rm c}$ with reciprocal lattice vectors $\boldsymbol{\tau}_{\rm c}=(\pm2,\pm2,0)$ and $(\pm2,\mp2,0)$ [See Fig.~\ref{fig:2}(a)]. Under $P\!=\!1.7~{\rm GPa}$ of pressure, these peaks do not noticeably change, indicating that the crystal mosaic remained intact, as expected given the large bulk modulus, $K_{0} \approx 190~{\rm GPa}$~\cite{Yong12:196}. No evidence of a pressure-induced structural phase transition was found in the paramagnetic phase [See Suppl. Fig.~\ref{fig:si-structure}], consistent with a previous study which found no phase transition below 14.2 GPa~\cite{Wang02:63}.

In the unpressurized sample, we observe magnetic Bragg peaks at ${\bf Q} = \boldsymbol{\tau}_{\rm c} \pm \boldsymbol{k}$ associated with two magnetic propagation vectors $\boldsymbol{k}_{\rm L,1}=(0.5,0.5,0)$ and $\boldsymbol{k}_{\rm L,2}=(0.5,0,1)$ {and the corresponding index permutations} [See Fig.\ref{fig:2}(a)-(b) left panels]. Also present are peaks consistent with equivalent domains generated by $90^{\circ}$ rotations about the $a$, $b$ and $c$ axes. Whilst not all of these operations are strictly symmetry operations of the $I4_{1}/amd$ space group, the tiny mismatch between the $a$ and $c$ lattice parameters $c/a= 0.998$~\cite{Ortega-San-Martin08:20} (in the cubic setting) means that structural domains of this type are energetically inexpensive, and similar domains have been observed in both MgCr$_{2}$O$_{4}$~\cite{Gao18:97} and other spinel compounds~\cite{Murakami13:577}. As a result, the neutron scattering data were symmetrized in the $m\overline{3}m$ Laue class, consistent with the point group of the crystal structure before the magneto-structural transition. Since the domain population is likely to vary as a function of various parameters (i.e. rate of compression, cooling, etc.), the symmetrization averages the domain populations along with improving counting statistics.  In addition to these magnetic propagation vectors, previously reported in Ref.~\onlinecite{Bai19:122}, peaks associated with $\boldsymbol{k}_{\rm H}=(0,0,1)$ are present but extremely weak under ambient pressure. 

\begin{figure*}
\centering
    \includegraphics[width=0.33\textwidth,trim=0cm 0cm 0cm 0cm,clip=true]{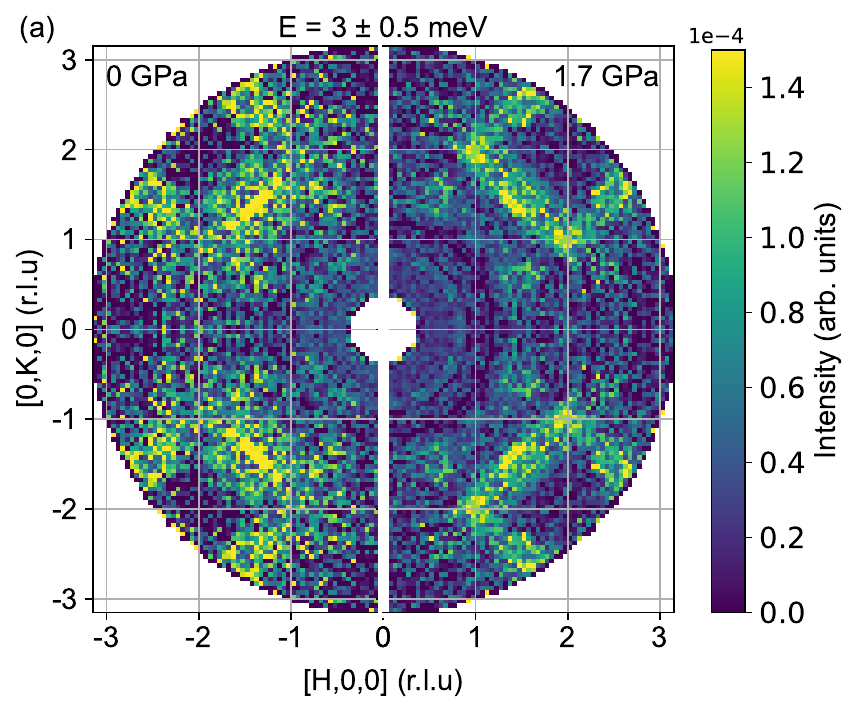}
    \includegraphics[width=0.33\textwidth,trim=0cm 0cm 0cm 0cm,clip=true]{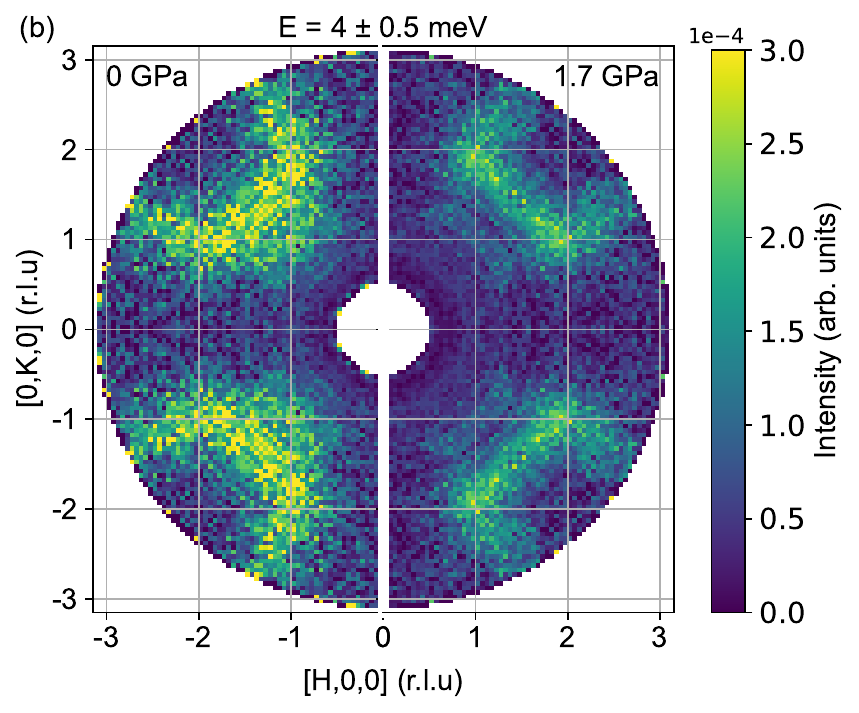}
    \includegraphics[width=0.33\textwidth,trim=0cm 0cm 0cm 0cm,clip=true]{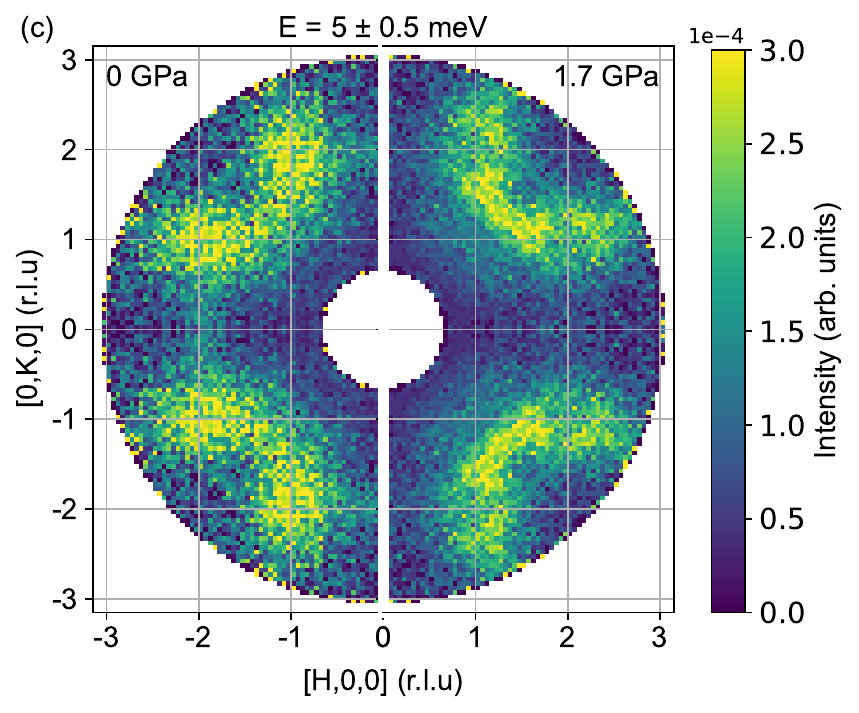}
    \caption{Momentum-dependence of the inelastic neutron scattering intensity $\tilde{I}({\bf Q},\bar{E}\textbf{})$ in the ordered phase at  $T\!=\!4.5$~K projected in the $(H,K,\bar{L})$ plane after integrating over $\bar{L} = 0 \pm 0.4$~r.l.u corresponding to the entire $\pm7.5^\circ$ out-of-plane detector coverage of HYSPEC and (a) $\bar{E}=3\pm 0.5$~meV, (b) $\bar{E}=4\pm 0.5$~meV, and (c) $\bar{E}=5\pm 0.5$~meV. Ambient pressure $P=0$~GPa is on the left while $P=1.7$~GPa is on the right.}
    \label{fig:4}
\end{figure*}

After the application of pressure, a redistribution of intensity is observed [See Fig.~\ref{fig:2}(a)-(b) right panels]. To quantify this change, in Fig.~\ref{fig:3} we turn to one-dimensional cuts through the elastic data, with integration ranges indicated by the white dashed boxes in Fig.~\ref{fig:2}. Figure \ref{fig:3}(a) shows a one-dimensional cut along $\bar{\bf Q}=(H,2,0)$. The Bragg peak at ${\bf Q} = (1,2,0)$, associated with $\tau_{\rm c}=(0,2,0)$ and $\boldsymbol{k}_{\rm H}=(1,0,0)$, dramatically increases in spectral weight by a factor of $\approx4.7$ between $P=0~{\rm GPa}$ and $1.7~{\rm GPa}$. Meanwhile, peaks associated with the propagation vectors $\boldsymbol{k}_{\rm L,1}$ and $\boldsymbol{k}_{\rm L,2}$ lose intensity with pressure. For instance, Fig.~\ref{fig:3}(b) shows the ${\bf Q} = (0.5,2,0.5)$ peak, associated with  $\tau_{\rm c}=(0,2,0)$ and $\boldsymbol{k}_{\rm L,1}=(0.5,0,0.5)$ decreasing in spectral weight by a factor of $\approx\!0.1$ compared to under ambient pressure. Likewise, Fig.~\ref{fig:3}(c) shows the ${\bf Q} = (1,1.5,0)$ [$\tau_{\rm c}=(0,2,0)$ and $\boldsymbol{k}_{\rm L,2}=(1,-{0.5},0)$] and ${\bf Q} = (2.5,1.5,0)$ [$\tau_{\rm c}=(2,2,0)$ and $\boldsymbol{k}_{\rm L,1}=(0.5,-{0.5},0)$] peaks decrease to factors of $\approx\!0.3$ and $\approx\!0.8$ from their ambient pressure spectral weight, respectively. Our elastic neutron scattering data thus demonstrates that applying GPa-scale hydrostatic pressure on MgCr$_2$O$_4$ leads to a spectral redistribution between the competing magnetic domains already present at ambient pressure.

\begin{figure}
\centering
    \includegraphics[width=0.99\columnwidth,trim=0cm 0cm 0cm 0cm,clip=true]{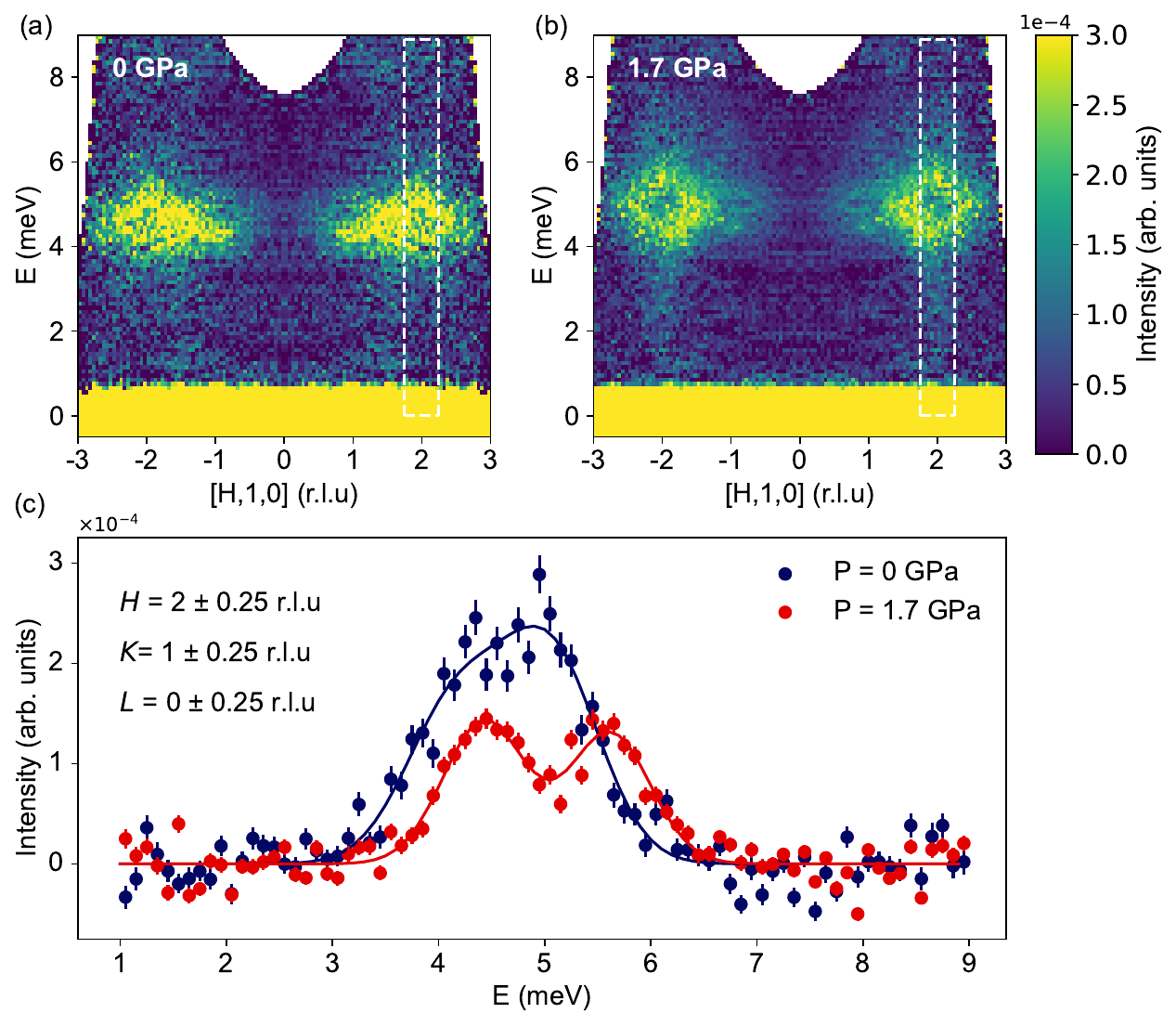}
    \caption{Momentum-energy dependence of the inelastic neutron scattering intensity $\tilde{I}(\bar{\bf Q},{E})$ in the ordered phase at $T\!=\!4.5$~K along the $\bar{{\bf Q}} = (H,1,0)$ direction with 
    $\bar{K}=1\pm0.25$~r.l.u and $\bar{L}=0\pm0.25$~r.l.u for (a) $P=0~{\rm GPa}$ and (b) $P=1.7~{\rm GPa}$. (c) Energy dependence of the inelastic neutron scattering intensity integrated around ${\bf Q} = (2,1,0)$ (integration ranges are indicated on the plot) as indicated by the dashed white lines in (a) and (b). Data have been symmetrized according to the $m\bar{3}m$ Laue group.}
    \label{fig:5}
\end{figure}

\begin{figure}
\centering
    \includegraphics[width=0.99\columnwidth,trim=0cm 0cm 0cm 0cm,clip=true]{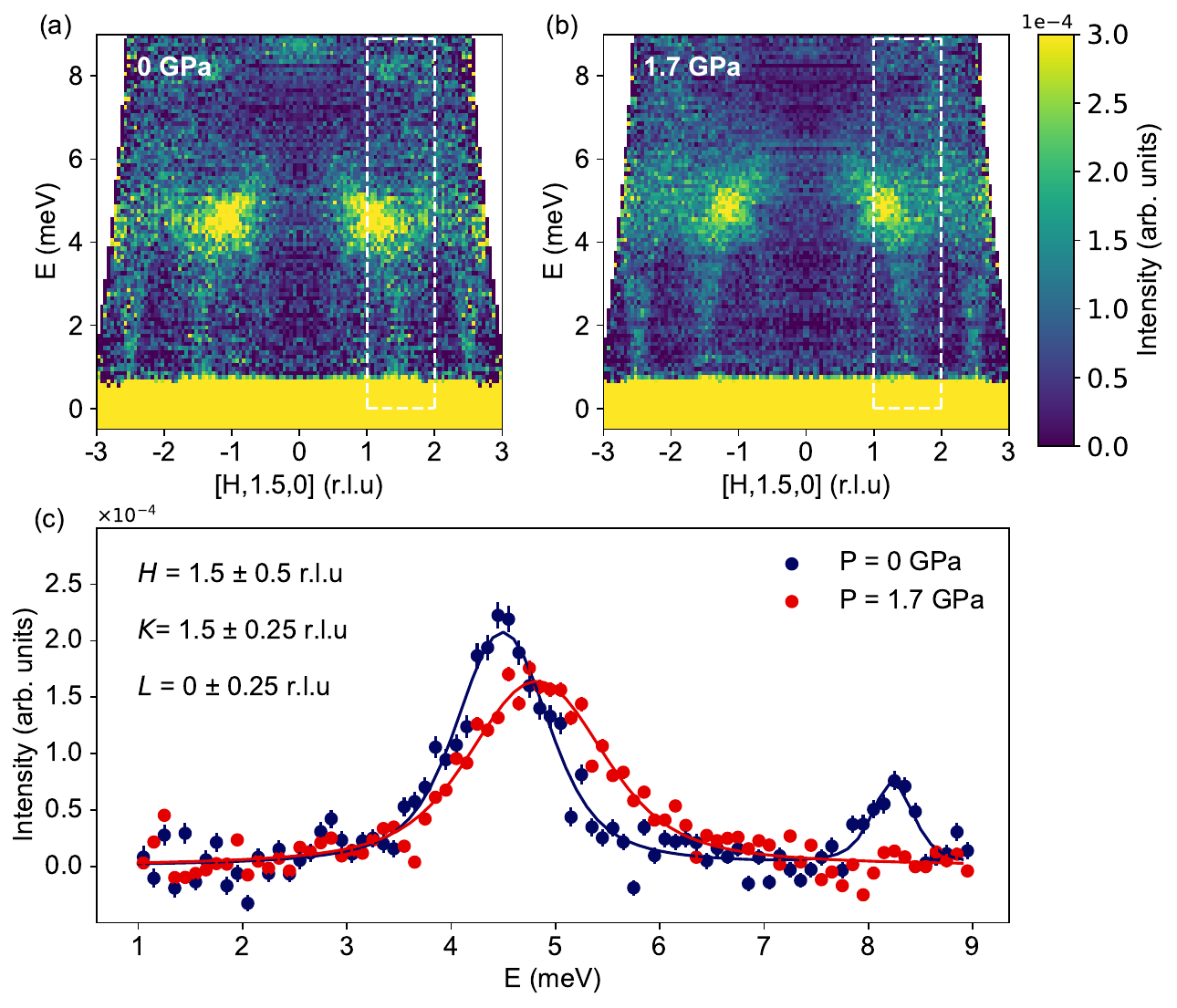}
    \caption{Momentum-energy dependence of the inelastic neutron scattering intensity $\tilde{I}(\bar{\bf Q},{E})$ at $T\!=\!4.5$~K along the $\bar{{\bf Q}} = (H,1.5,0)$ direction with $K=1.5\pm0.25$~r.l.u and $K=0\pm0.25$~r.l.u for (a) $P=0~{\rm GPa}$ and (b) $P=1.7~{\rm GPa}$. (c) Energy dependence of the inelastic neutron scattering intensity integrated around ${\bf Q} = (1.5,1.5,0)$ (integration ranges are indicated on the plot) as indicated by the dashed white lines in (a) and (b). Data have been symmetrized according to the $m\bar{3}m$ Laue group.}
    \label{fig:6}
\end{figure}

\begin{figure}
\centering
    \includegraphics[width=0.99\columnwidth,trim=0cm 0cm 0cm 0cm,clip=true]{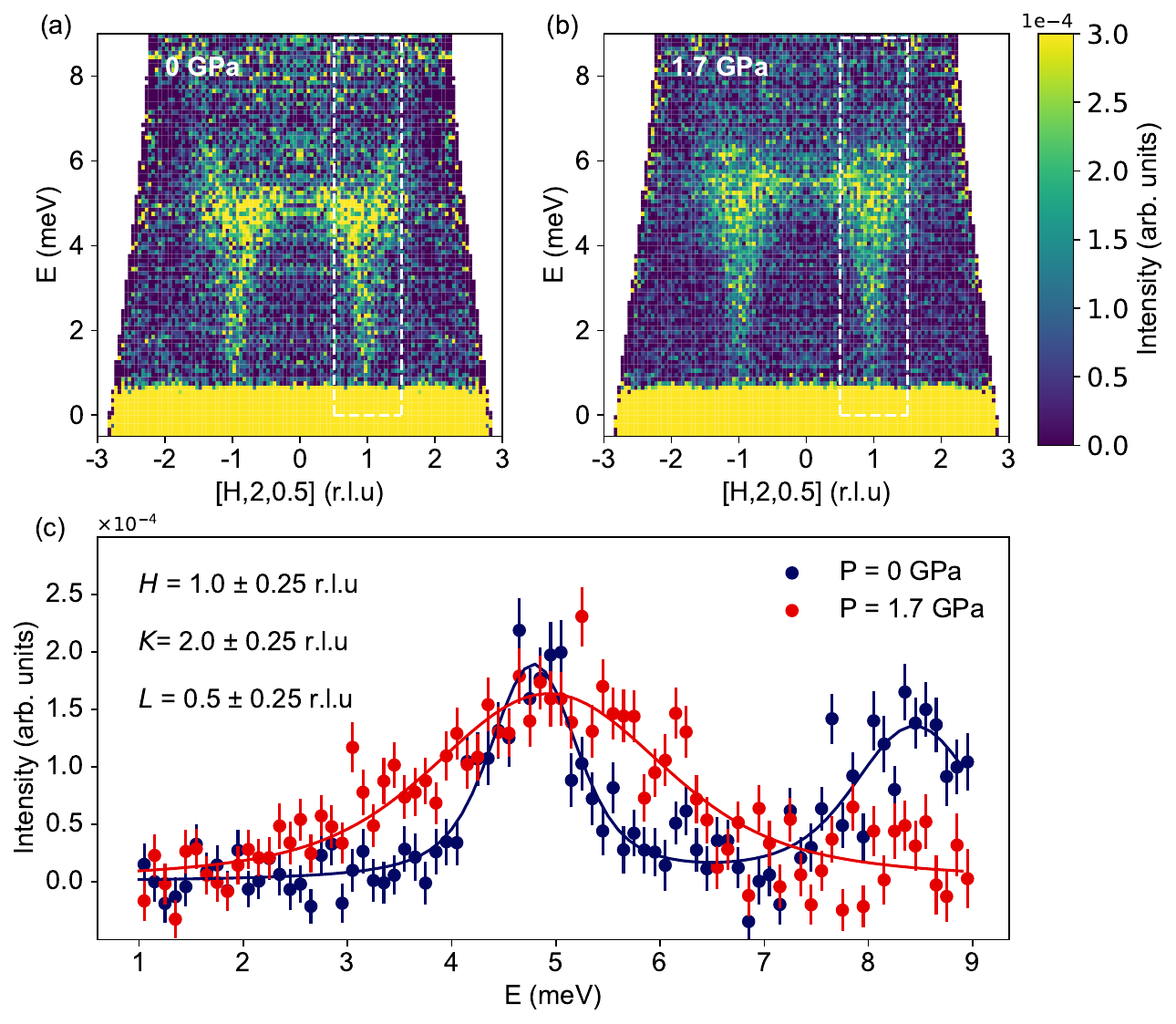}
    \caption{Momentum-energy dependence of the inelastic neutron scattering intensity $\tilde{I}(\bar{\bf Q},{E})$ at $T\!=\!4.5$~K along the $\bar{{\bf Q}} = (H,2,0.5)$ direction with $K=1.5\pm0.25$~r.l.u and $K=0\pm0.25$~r.l.u for (a) $P=0~{\rm GPa}$ and (b) $P=1.7~{\rm GPa}$. (c) Energy dependence of the inelastic neutron scattering intensity integrated around ${\bf Q} = (1.5,1.5,0)$ (integration ranges are indicated on the plot) as indicated by the dashed white lines in (a) and (b). Data have been symmetrized according to the $m\bar{3}m$ Laue group.}
    \label{fig:6p5}
\end{figure}

\subsubsection{Inelastic Scattering}

The intensity redistribution between magnetic Bragg peaks under pressure is visibly reflected in the spectral structure of the magnetic excitations. Constant-energy slices through the $T=4.5~{\rm K}$ data projected in the $(H,K,\bar{L}=0)$ plane give a first hint at that evolution [See Fig.~\ref{fig:4}]. The magnetic excitation spectrum of MgCr$_2$O$_4$ in the ordered phase is dominated by a concentration of spectral weight between $E\approx4.0$ and $5.5$~meV~\cite{Tomiyasu08:101,Gao18:97}. Cuts spanning that resonance-like band reveal a rounded structure factor under ambient pressure that becomes more squared up at $P=1.7~{\rm GPa}$ [Fig.~\ref{fig:4}(a)-(c)], while somewhat losing intensity at the lowest two energies. As constant-energy slices reflect the spatial dependence of the two-point spin correlations, we directly associate this change with the evolution of the magnetic structure. Indeed, a close inspection at the lowest energy [Fig.~\ref{fig:4}(a)], shows spectral weight shifted towards ${\bf Q} = (1,2,0)$ and symmetrically equivalent positions, for which underlying Bragg peaks are enhanced under pressure. This behavior suggests that the band of excitations between $E\approx4.0$ and $5.5$~meV originates from overlapping spin-wave excitations associated with the different coexisting Bragg peaks.

The effect of applied pressure is even more apparent in the momentum-energy dependence of the inelastic scattering along reciprocal lattice directions crossing magnetic Bragg peak positions [Fig.~\ref{fig:5}, Fig.~\ref{fig:6} and Fig.~\ref{fig:6p5}]. The response along $\bar{\bf Q} = (H,1,0)$ at $P=1.7~{\rm GPa}$ [Fig. \ref{fig:5}(b)] reveals acoustic spin wave modes emerging from the ${\bf Q}=(2,1,0)$ Bragg peaks associated with the $\boldsymbol{k}_{\rm H}$ magnetic propagation vector. These excitations are weaker in the $P=0~{\rm GPa}$ data [Fig. \ref{fig:5}(a)]. Additionally, focusing on the energy response at Bragg peak position ${\bf Q}=(2,1,0)$ shows that the dominant excitation shifts in energy by $\approx 0.5$~meV and splits into two distinct excitations at $E\approx4.5$~meV and $E\approx5.5$~meV [Fig. \ref{fig:5}(c)]. The energy FWHM of these excitations of $\Gamma \approx 0.8(1)$~meV goes beyond the expected spectrometer's resolution of $0.34$~meV, but the momentum-integration may be at the origin of this broadening. The enhancement of acoustic spin-waves under pressure is also apparent, although less dramatic, along $\bar{\bf Q} = (H,1.5,0)$ at $P=1.7~{\rm GPa}$ [Fig. \ref{fig:6}(b)] compared to $P=0~{\rm GPa}$ [Fig. \ref{fig:6}(a)]. This is also accompanied by an $\approx 0.5$~meV energy shift of the lower-energy excitation band for ${\bf Q}=(1.5,1.5,0)$ [Fig. \ref{fig:6}(c)], as well as the disappearance of higher energy excitations observed around $E\approx8$~meV for $P=0~{\rm GPa}$. This excitations are presumably pushed to higher energies and outside the kinematic constraints of our $E_i=12$~meV data, althought our $E_i=25$~meV is inconclusive. Overall, this shows that the intensity of the resonance-like spectral features in the ordered phase of MgCr$_2$O$_4$ are correlated with the magnetic propagation vectors of the underlying magnetic ordering. Furthermore, it evidences that a pressure of $P=1.7~{\rm GPa}$ stiffens excitations by at least $0.5$~meV, corresponding to $\delta E \approx 0.25~{\rm meV}$ per GPa ($\delta E/E_{\rm mode} \approx 10 \%$), as judged by the global upward shift of the spectrum. The response along $\bar{\bf Q} = (H,2,0.5)$ [Fig. \ref{fig:6p5}] shows qualitatively similar behavior to that along $\bar{\bf Q} = (H,0.5,0)$. Dispersive spin wave excitations are seen emerging from the $\mathbf{Q}=(1,2,0.5)$ Bragg peaks both under ambient [Fig \ref{fig:6p5}(a)] and applied [Fig \ref{fig:6p5}(b)] pressure. The disappearance of higher energy excitations is again seen [Fig \ref{fig:6p5}(c)]. The qualitative similarities between Fig. \ref{fig:6} and Fig. \ref{fig:6p5} result from structural domains. It should be noted this is not an artifact of symmetrization since the data have only been symmetrized in the scattering plane.

\section{Discussion and Conclusion}
\label{Sect:Discussion}

Overall, the application of GPa-scale pressure on the magnetism of polycrystalline and single-crystaline samples of MgCr$_2$O$_4$ manifests in three interrelated but subtle effects: the enhancement of the N{\'e}el ordering transition by around $\approx0.8~{\rm K}$ per GPa, a large redistribution in the population of the three nearly degenerate magnetic ordering wave-vectors which coexist in the ordered state, and an increase in the bandwidth of the resonance-like excitations by around $\approx 0.25~{\rm meV}$ per GPa. Below, we discuss possible mechanisms for these results and the implications for our understanding of the MgCr$_2$O$_4$'s ordered state.

\subsection{Exchange interactions and pressure}

The interplay between the direct $d$-$d$ and the oxygen-mediated superexchange between Cr$^{3+}$ ions is expected to be important in MgCr$_2$O$_4$, in which the magnetic ions occupy a network of edge-sharing octahedral with 90$^{\circ}$ Cr-O-Cr superexchange pathways. The half-filling of the $t_{2g}$ orbitals makes the $d_{xy}$, $d_{xz}$ and $d_{yz}$ orbitals ``active'' in the superexchange sense. Direct $d$-$d$ exchange in the Cr$^{3+}$ spinels favors antiferromagnetic coupling while the oxygen-mediated exchange instead favors a weakly ferromagnetic coupling since $d$-orbitals on neighboring sites overlap with orthogonal $p$-orbitals~\cite{Khomskii:book}. The antiferromagnetic correlations present in MgCr$_{2}$O$_{4}$ are suggestive of dominant $d$-$d$ overlap and hence the application of pressure might be expected to suppress the magnetic order as the nearest neighbor AFM bond gains an enhancement from $d$-$d$ exchange yet the longer range further neighbor interactions do not. 

However, the presence of strong magnetoelastic coupling in the spinel oxides complicates matters further. Indeed in the Cd, Hg and Zn chromium spinels, the magnetostructural transition has been observed to increase in temperature~\cite{Ueda08:77}. This increase in the ordering temperature may be understood in the context of the bond phonon model. Expanding the exchange interaction in small displacement, one finds that the linear reduction in magnetic exchange energy wins over the quadratic energy cost of harmonic oscillation~\cite{Tchernyshyov02:88} destabilizing the system towards a spin-Peierls state. The magnetoelastic energy is $E\sim -f^{2}(\partial J/\partial x)^{2}/2k$, where $k$ is the spring constant and $\boldsymbol{f}$ is a bond order parameter~\cite{Tchernyshyov02:88}. The enhancement of the short range $d$-$d$ exchange can therefore be expected to give rise to an enhanced magnetoelastic coupling since the direct exchange strength, to a first approximation, depends exponentially on distance~\cite{Freeman61:124}. Controlling the ratio of $d$-$d$ to $d$-$p$-$d$ exchange through hydrostatic pressure may thus yield control of the magnetostructural coupling.

\subsection{Magnetic ordering}

The thermal hysteresis and the observation of two distinct peaks in both the susceptibility and specific heat of MgCr$_2$O$_4$ may indicate that two distinct magnetic ordering wave-vectors are associated with the N{\'e}el transition at $T_{\rm N}$. This is in agreement with the evidence presented in Ref.~\onlinecite{Gao18:97} which strongly suggests that the magnetic ordering in $\mathrm{MgCr_{2}O_{4}}$ is multi-domain rather than multi-$\boldsymbol{k}$. Additionally, the emergence of a peak in  $\mathrm{\partial(\chi\cdot T)/\partial T}$ between $T=15.0$ and $16.0$~K may correspond to the transition between the paramagnetic and H-phase. A signature of this transition in susceptibility data has yet to be reported; indeed, it is absent in our data collected under ambient pressure. Given that elastic neutron scattering results at $P\!=\!1.7~{\rm GPa}$ demonstrate that the population of the $\boldsymbol{\kappa}_{\rm H} = (1,0,0)$ magnetic domain increases by almost a factor 5 compared to ambient pressure, we interpret the appearance of this peak in our thermomagnetic measurements as possible evidence of the transition into the H-phase. 

A complete magnetic and structural description of the low-temperature phase of MgCr$_2$O$_4$ has thus far evaded elucidation, despite comprehensive magnetic structural studies in Ref. \onlinecite{Gao18:97} using neutron diffraction and spherical neutron polarimetry. A unique magnetic structure solution was found for the $\boldsymbol{k}_{L,2}$ domain, but two indistinguishable solutions were found for $\boldsymbol{k}_{L,1}$. It is, therefore, impossible to determine whether the two low-temperature structures are strictly degenerate or lie close in energy. One possible explanation for the presence of co-existing ordering wave-vectors is the existence of stacking faults~\cite{Lin84:20} or antiphase domain boundaries~\cite{Fogh17:96,Lane21:104,Edwards20:102} pinned by structural boundaries or defects. It is interesting to consider whether a detwinned crystal would show a similar phase coexistence. Moreover, a magnetic structural refinement of such a crystal would allow a unique determination of the $\boldsymbol{k}_{L,1}$ due to anisotropy inherent to one of the two proposed models~\cite{Gao18:97}.

\subsection{Magnetic fluctuations}
Previous studies have suggested that the excitations in the Cr$^{3+}$ spinels are consistent with decoupled local spin clusters, with resonances associated with hexamers and tetramers~\cite{Lee02:418,Tomiyasu08:101,Gao18:97}, while other studies~\cite{Conlon10:81}, including by some of us~\cite{Bai19:122}, strongly favor an interpretation in terms of competing further neighbor exchange interactions for a Heisenberg model. The argument for the hexagonal spin-loop cluster as the relevant low-energy degree of freedom stemmed both from the nature of the observed magnetic fluctuations and the principle that the ground state of the nearest-neighbor pyrochlore lattice Heisenberg antiferromagnet satisfies the zero-divergence condition $\sum_{t} \sum_{i=1}^4 {\bf S}_{t,i}=0$, where $t$ runs over the lattice tetrahedrals. Since each side of a hexagon links two spins on a tetrahedron, for a collinear antiferromagnetic arrangement of spins on the hexagon, there is a local zero energy mode -- the staggered magnetization of the loop can be rotated at no energy cost since their neighbor exactly cancels their contribution to the total spin on each tetrahedron. These zero energy modes are not unique to the clusters but can take the form of longer-length strings of collinear spins~\cite{Tchernyshyov02:88}. It seems \textit{a priori} reasonable that the strings of the shortest length, the hexagonal cluster, might be preferred in the presence of finite string tension. The presence of finite string tension is suggested in the low-temperature phase by the lifting of the zero mode to finite energy but is also present above $T_{\rm N}$ due to further-neighbor exchange interactions~\cite{Bai19:122}.  

\begin{figure}
	\centering
	    \includegraphics[width=\linewidth,trim=0cm 0cm 0cm 0cm,clip=true]{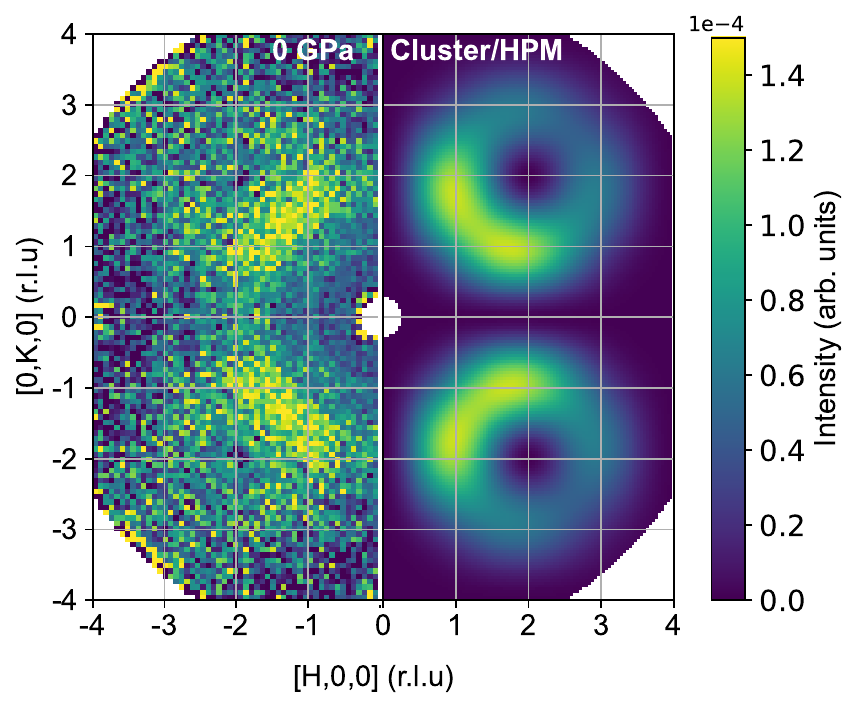}
	    \caption{Momentum-dependence of the inelastic neutron scattering intensity $\tilde{I}({\bf Q},\bar{E}\textbf{})$ measured with $E_i=25$~meV in the paramagnetic phase at $T\!=\!20$~K projected in the $(H,K,\bar{L})$ plane after integrating over $\bar{L} = 0 \pm 0.5$~r.l.u and $\bar{E}=4\pm 2$~meV to minimize background signal. Comparison with calculation for hexagonal clusters.}
	    \label{fig:7}
	\end{figure}

In essence, the dynamical spin structure factor of putative localized clusters exhibits a decoupling between energy and momentum dependence, $\mathcal{S}(\mathbf{Q},E)=\sum_r S_r(\mathbf{Q}) g(E)$ where $g(E)=\sum_r \delta(E-E_{r})$ and $E_{r}$ are the resonance frequencies of the clusters. The momentum dependence merely stems from the form factor associated with the spatial extent of a given spin cluster, $S_r(\mathbf{Q})=|F_r(\mathbf{Q})|^{2}$, which has a well-known closed form for the various cluster shapes~\cite{Tomiyasu08:101,Tomiyasu13:110,Bai19:122,Lee02:418}. This functional form reproduces the energy-integrated structure factor $\mathcal{S}(\mathbf{Q})=\int d E \mathcal{S}(\mathbf{Q},E)$, however, the presence of dispersive excitations [Figs.~\ref{fig:5}, \ref{fig:6}, and \ref{fig:7}, as well as Ref.~\cite{Gao18:97}] suggests wave-like excitations in MgCr$_{2}$O$_{4}$ extending over a length scale greater than that of a single cluster. Further, in Ref.~\onlinecite{Bai19:122} it was demonstrated that the dynamics above the magnetic ordering transition could be understood as coherent spin precessions (i.e., linear spin wave modes) riding a disordered background.  

The apparent similarity between $\mathcal{S}(\mathbf{Q})$ for a Heisenberg paramagnet and the cluster model stems from the similarity between the structure factor of the hexagonal spin cluster model $\mathcal{S}_{\rm cluster}(\mathbf{Q})=\frac{2}{3} \sum_{i=0}^{3}\langle \mathbf{S}_{0} \cdot \mathbf{S}_{i}\rangle \mathrm{cos}(\mathbf{Q}\cdot\mathbf{r}_{0i})$ compared to a Heisenberg paramagnet~\cite{Bai19:122} $\mathcal{S}_{\rm HPM}(\mathbf{Q})=\frac{2}{3N} \sum_{ij}\langle \mathbf{S}_{i} \cdot \mathbf{S}_{j}\rangle \mathrm{cos}(\mathbf{Q}\cdot\mathbf{r}_{ij})$. Summing the cluster structure factor over all hexagonal tilings of the pyrochlore lattice captures all bonds up to the third nearest neighbor such that their Fourier weights are identical to those of the Heisenberg paramagnet. It is thus unsurprising that self-consistent Gaussian approximation (SCGA) calculations for a Heisenberg Hamiltonian with further neighbor exchanges strongly agree with the cluster calculation~\cite{Bai19:122} [See Fig.\ref{fig:7} for a demonstration of that effect]. Similarly, for the average over tetragonal clusters, the Fourier functions have the same functional form as a Heisenberg paramagnet.

In this context, our experiments under pressure shine an essential light on the nature of the excitations in the ordered phase of MgCr$_2$O$_4$. The dispersive nature of the low energy excitations, including the apparent resonance $E \approx 4.5$ meV, is more clearly visible under pressure [See Figs.~\ref{fig:5} and \ref{fig:6}]. This shows that the neutron scattering intensity previously described as a spin cluster resonance undoubtedly results from overlapping spin wave branches poorly separated in momentum and energy. These originate from the enlargement of the magnetic unit cell compared to the chemical unit cell and the coexistence of magnetic domains. These modes are difficult to resolve at ambient pressure, however by subtly modifying the competing energy scales in the system, we have reduced the occupation of the $\boldsymbol{\kappa}_{L,2} = (0.5,0,1)$ magnetic domain allowing for the two spin wave branches to be resolved [See Fig. \ref{fig:5} (c)].

\subsection{Conclusion}

We have presented thermo-magnetometic and neutron scattering measurements on polycrystalline and single-crystalline samples of MgCr$_{2}$O$_{4}$ in applied, pseudo-hydrostatic, GPa-scale pressures. Our magnetometry measurements at ambient conditions are consistent with previous measurements of stoichiometric MgCr$_{2}$O$_{4}$ however we are able to resolve two peaks at the magnetostructural transition in our field-cooled data, adding further weight to the argument that the low-temperature phase of this material is multi-domain rather than multi-$\boldsymbol{k}$. An increase in $T_{N}$ under pressure is suggestive of an increased direct exchange and a subtle modification of the competing energy scales in the system. Our neutron measurements under ambient conditions replicate those of previous studies~\cite{Bai19:122,Gao18:97} however, under pressure of $P\!=\!1.7$~GPa we see a decrease in the population of one of the magnetic domains of the L-phase $\boldsymbol{k}_{L,2}$ and a gain in population of the H-phase $\boldsymbol{k}_{H}$. The shift in the energy of magnetic excitations and the redistribution of spectral weight to spin waves emanating from different Bragg peaks allows for the overlapping spin wave branches at $E\approx 4.5$ meV to be resolved, unambiguously revealing coherent spin wave excitations rather than localized cluster modes.


\begin{acknowledgements}
We are grateful to Sîan Dutton for her insights on interpreting our susceptibility data and to Collin Broholm for inspiration at the inception of this project. This project at GT (L.N., H.L., M.M.) was funded by the U.S. Department of Energy, Office of Basic Energy Sciences, Materials Sciences and Engineering Division under Award DE-SC-0018660. H. L. acknowledges additional financial support from the Royal Commission for the Exhibition of 1851. This research used resources at the High Flux Isotope Reactor and Spallation Neutron Source, a DOE Office of Science User Facility operated by the Oak Ridge National Laboratory. The single-crystal synthesis work was supported as part of the Institute for Quantum Matter, an Energy Frontier Research Center funded by the U.S. Department of Energy, Office of Science, Basic Energy Sciences under Award DE-SC0019331. This work utilized the Materials Characterization Facility (MCF) jointly supported by Georgia Tech's Institute for Materials (IMat) and the Institute for Electronics and Nanotechnology (IEN), which is a member of the National Nanotechnology Coordinated Infrastructure supported by the National Science Foundation under Award ECCS-2025462.
\end{acknowledgements}

\appendix 

\section{Hydrostaticity and peak width in $\mathrm{\partial(\chi\cdot T)/\partial T}$}
\label{sec:appendix-hydrostatic}

In this appendix, we explore how the width of the peak in $\partial [\chi\!\cdot\!T]/\partial T\vert_H$, extracted from our magnetometry measurements, correlates with experimental conditions and may be used to test the hydrostatic pressure conditions experienced by the sample. Assuming that each external pressure maps onto a single and well-defined magnetic transition temperature, it appears possible in principle to correlate the width of the peak in $\partial [\chi\!\cdot\!T]/\partial T\vert_H$ to pressure gradients and heterogeneities. For instance, under hydrostatic pressure conditions, the bulk of the sample should display a single transition temperature with a narrow peak in  $\partial [\chi\!\cdot\!T]/\partial T\vert_H$. Conversely, the lack of hydrostaticity should translate into a distribution of transition temperatures and a subsequently wider peak. 

	\begin{figure}[b!]
	\begin{center}
	    \includegraphics[width=0.9\linewidth,trim=0cm 0cm 0cm 0cm,clip=true]{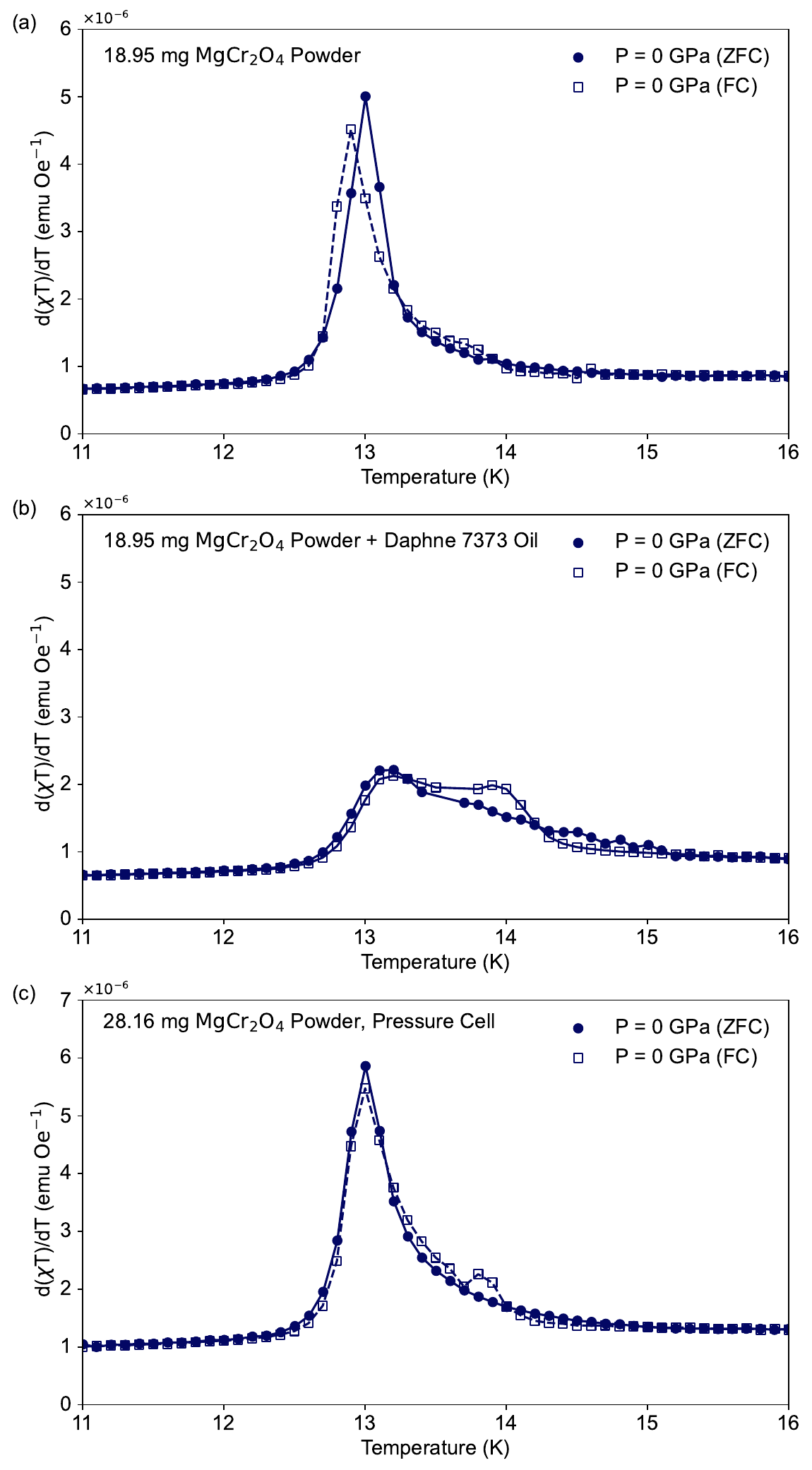}
	    \end{center}
	    \caption{ (a) $\partial [\chi\!\cdot\!T]/\partial T\vert_H$ at $\mu_0H=0.5$~T for a control sample consisting of MgCr$_2$O$_4$ in a standard VSM capsule and brass sample holder. (b) The same sample was combined with daphne oil in a standard VSM capsule, (c) the same sample was loaded into the pressure cell with no daphne oil.}
	    \label{fig:A1}
	\end{figure}

A major challenge to this approach, however, is that the limiting width of the peak in $\partial [\chi\!\cdot\!T]/\partial T\vert_H$ may depend on sample quality, thermalization, and other extrinsic effects. Indeed, the peak we observe in our samples [See Fig.~\ref{fig:1}] is much broader than that of the reference stoichiometric sample~\cite{Dutton11:83}. We conducted several tests of sample quality and thermalization to understand if these effects play a role. First, we measured our sample outside the pressure cell in a VSM sample holder, using the same measurement sequence employed in previous measurements. The peak derivative appears much narrower [see Fig.~\ref{fig:A1}(a)] than in the pressure cell [See Fig.~\ref{fig:1}] and in agreement with the literature report~\cite{Dutton11:83}. Next, Pb powder was added to the same VSM capsule [Fig. not shown] with no visible change. Third, the Daphne 7373 oil, used in our experiments as PTM, was added to the capsule containing the sample and Pb, until both powders were saturated with oil. This lead to a considerable broadening of the peak in $\partial [\chi\!\cdot\!T]/\partial T\vert_H$ [see Fig.~\ref{fig:A1}(b)]. Finally, a sample of MgCr$_2$O$_4$ from the same batch was measured inside the pressure cell but without Pb and Daphne 7373 oil. In that case, we recovered a sharp peak with only minor broadening [see Fig.~\ref{fig:A1}(c)]. 

\begin{figure}
	\begin{center}
	    \includegraphics[width=0.9\linewidth,trim=0cm 0cm 0cm 0cm,clip=true]{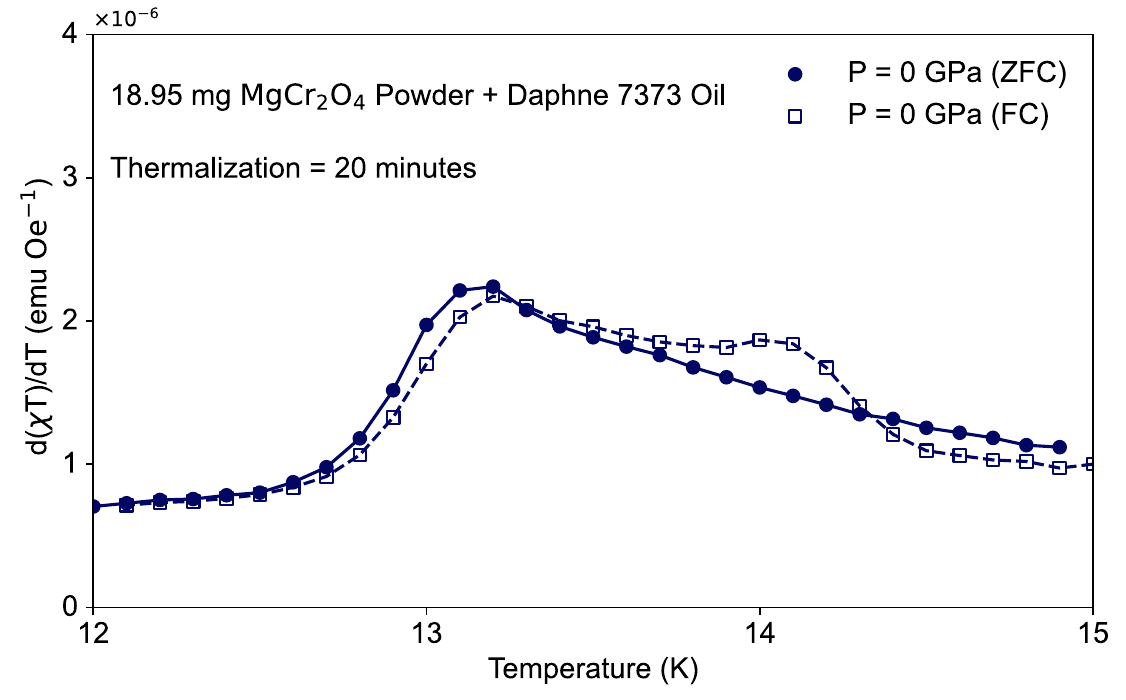}
	    \end{center}
	    \caption{$\mathrm{\partial(\chi\cdot T)/\partial T}$ at $\mu_0H=0.5$~T for the control sample combined with daphne oil and Pb measured after waiting 20 minutes for sample thermalization between the acquisition of each data point.}
	    \label{fig:A2}
	\end{figure}

Having identified the presence of Daphne 7373 oil as the dominant source of the broadening, we investigated if thermalization of the powder sample within the oil was a concern. To do so, we repeated measurements with a capsule containing the MgCr$_2$O$_4$ powder, Pb, and Daphne 7373 oil, but we increased the thermalization period between each data point in the sequence from a few seconds to 20 minutes. This did not yield any changes [see Fig.~\ref{fig:A2}] compared to the typical measurement sequence. We conclude that the broadening of the  $\partial [\chi\!\cdot\!T]/\partial T\vert_H$ peak in our data is a systematic effect of the Daphne 7373 oil, possibly related to the freezing transition of the oil at $T\approx200$~K \cite{Yokogawa07:46} imprinting an inhomogeneous stress field on the sample.

Overall, the baseline broadening due to the PTM may enhance a subtle shoulder in the FC measurements that is already present in the data without the Daphne 7373 oil. However, due to this inherent broadening, we consider the results of Fig.~\ref{fig:A1}(b) as the baseline peak width to which we compare those for compressions $\Delta \ell > 0$ when analyzing the hydrostaticity of our applied pressure. Since no significant peak broadening occurs from this initial reference, we have deemed our measurements sufficiently hydrostatic for the purposes of this report.

%

\clearpage
\widetext
\setcounter{equation}{0}
\setcounter{figure}{0}
\setcounter{table}{0}
\setcounter{page}{1}
\renewcommand{\theequation}{S\arabic{equation}}
\renewcommand{\thefigure}{S\arabic{figure}}
\begin{center}
\textbf{\large Supplemental Information}
\end{center}

\noindent{\bf 1. Phase purity of synthesized samples from Powder X-ray Diffraction}\\

\begin{figure}[h!]
    \centering
    \includegraphics[width = 0.80\linewidth]{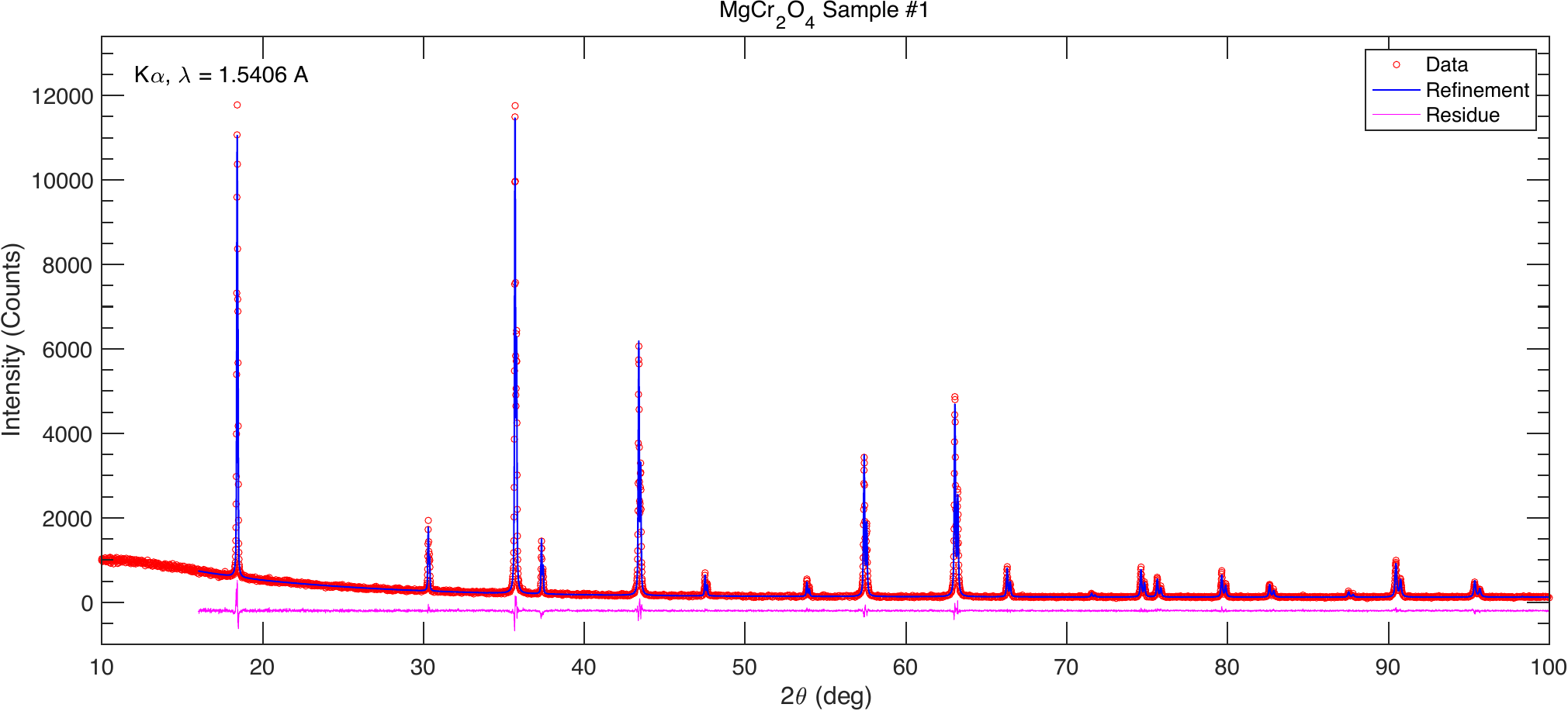}\\ \vspace{1cm}
    \includegraphics[width = 0.80\linewidth]{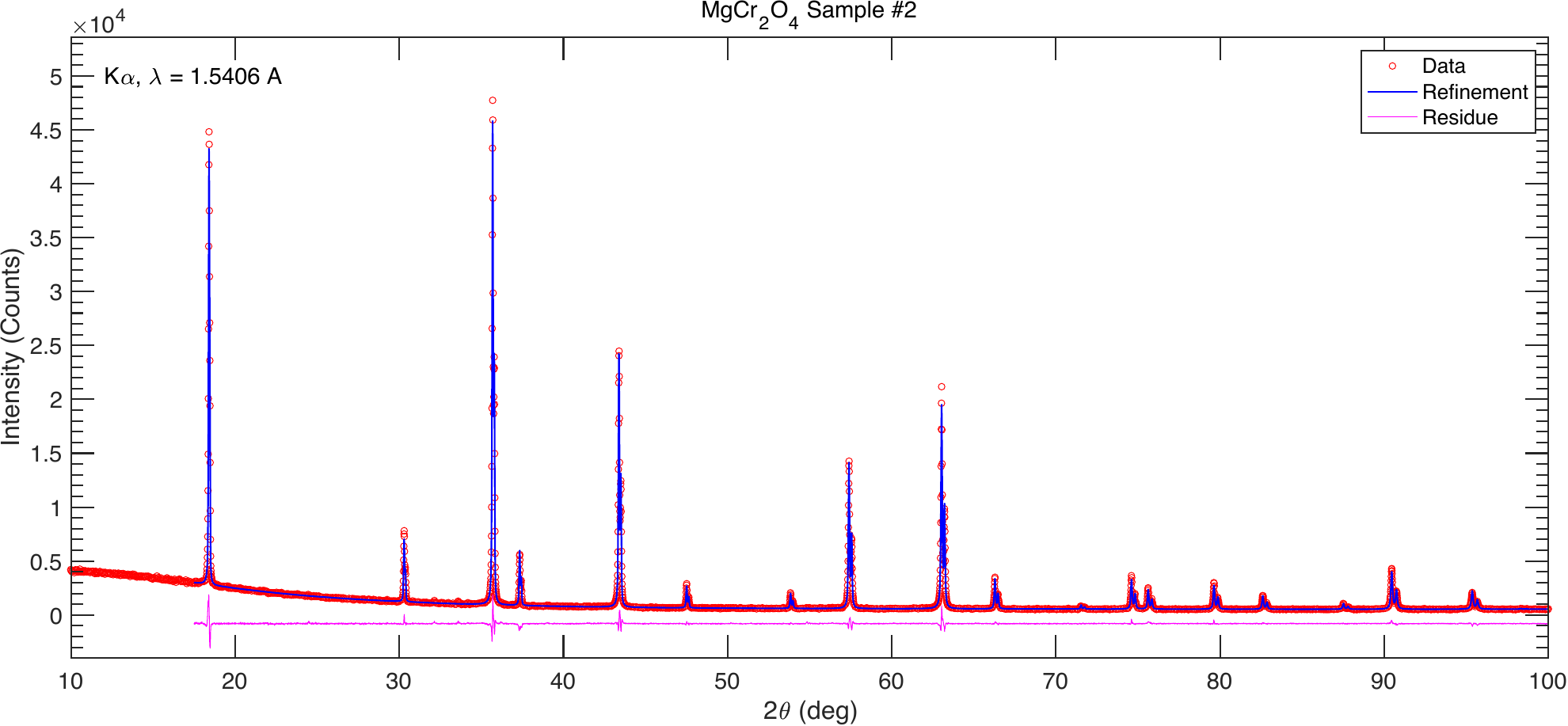}
    \caption{Powder x-ray diffraction pattern and corresponding Rietveld Refinements for (a) sample $\#1$ (refined $a=8.332656(16)$ \AA) and (b) sample $\#2$ (refined $a=8.333138(16)$\AA) grown in separate batches but under synthesis similar protocol.}
    \label{fig:si-pxrd}
\end{figure}

This study uses two distinct polycrystalline samples of MgCr$_2$O$_4$. Sample $\#1$ is studied in the main text, and sample $\#2$ is presented as an alternate. These samples come from batches that are synthesized according to the same protocol but prepared separately. Results of PXRD and Rietveld refinement using the Rigaku Smartlab II studio software are presented in \ref{fig:si-pxrd} for these two samples.

\pagebreak
\noindent{\bf 2. Pressure determination from Pb superconducting transition} \\

\begin{figure}[h!]
    \centering
    \includegraphics[width = 1.0\linewidth]{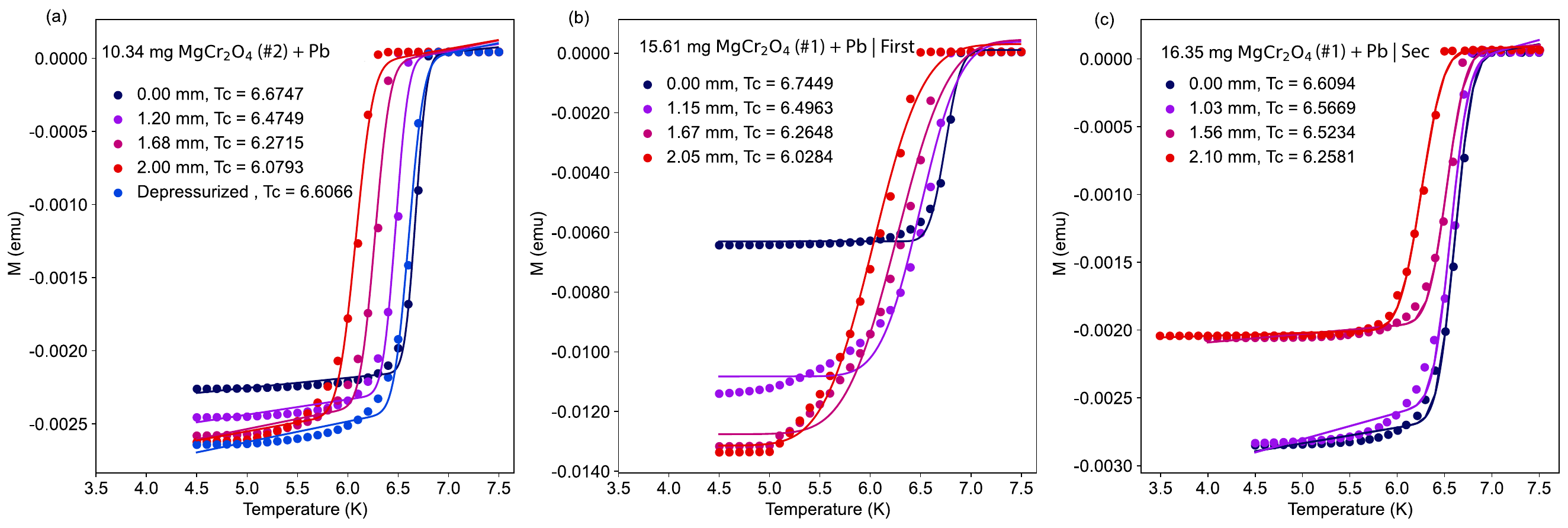}
    \caption{Onset of perfect diamagnetism in Pb powder mixed with a sample of MgCr$_2$O$_4$ and measured with $\mu_0H=0.01$~T for four different cell compressions $\Delta \ell$. The three different plots correspond to replicated experiments reaching different maximal applied pressure. (a) Measurement performed on sample $\#2$ (results not shown in the main text). (b) First measurement performed on sample $\#1$ (results not shown in the main text). (c) Second measurement performed on sample $\#1$ (shown in the main text in Fig.~\ref{fig:1}).}
    \label{fig:si-pbtc}
\end{figure}

The results of our in-situ pressure determination from the superconducting transition of Pb are shown in Fig.~\ref{fig:si-pbtc} with  Fig.~\ref{fig:si-pbtc}(c) corresponding to the measurements shown in  Fig.~\ref{fig:1} of the main paper. Below, we discuss some of our observations and experimental approaches, as these details may be interesting to other researchers in the field (or ourselves in the future). \\

Pb was deemed a suitable low-temperature manometer given that its pressure dependence is well characterized as a function of field and pressure. Often this relationship is reported at 0 Oe, however since applied fields of less than 10 Oe (0.001 T) are unreliable on our MPMS3 system (not equipped with an ultra-low-field option) without consistent resetting of the magnet, we measured the superconducting transition at $\mu_0H=100$~Oe (0.01 T) which still allowed the diamagnetic response of Pb below $T_c$ to dominate over the paramagnetic signal of our sample. Therefore, $T_c$ of Pb in our measurements needs to be determined by going beyond the usual linear equation of state,
    \begin{equation}
        T_{c}(P)=T_{c}(0)-m_{T}P,
    \end{equation}
where $m_{T}$ is experimentally determined~\cite{Eiling81:11}. Given that the critical temperature is expected to have both a field and pressure dependence, following Ref. \onlinecite{Suresh07:75}, we express the critical field as an even polynomial in $T$ and separate the temperature and pressure dependence~\cite{Suresh07:75} employing the ``similitude principle"~\cite{Suresh07:75,Garfinkel61:122}
\begin{equation}
    H_{c}(T,P)=H_{c0}(P)\left[1-\alpha \left(\frac{T}{T_{c}}\right)^{2}+(1-\alpha) \left(\frac{T}{T_{c}}\right)^{4}\right].
    \label{eq:similitude}
\end{equation}
\noindent In previous work, the zero-temperature pressure dependence has been fitted to~\cite{Suresh07:75} 
\begin{equation}
    H_{c0}(P)=H_{c0}(0)-m_{H}P.
\end{equation}
\noindent Substituting both the zero field critical temperature, $T_{c}(P)$ and the zero temperature critical field $H_{c0}(P)$ into Eqn. \ref{eq:similitude}, we can solve for the transition temperature as a function of pressure at fixed applied field, $H_{c}=H$. Using values of $T_{c}(0)=7.20$ K,  $m_{T}=0.365$, from Ref.~\onlinecite{Eiling81:11} and $\alpha=0.954$, $H_{c0}=803.72$ Oe and $m_{H}=67.028$ from Ref.~\onlinecite{Suresh07:75} we find that an estimate to the field-adjusted transition temperature at $H=100$ Oe is given by
\begin{equation}
    T_{c}^{100\mathrm{Oe}}(P)=6.76 {\rm K}-0.381P {\rm K/GPa}
    \label{eq:pressure-corrected}
\end{equation}
\noindent where we have neglected small nonlinear corrections in $P$. We therefore find a small ($\approx 4$\%) correction to $m_{T}$ in field. \\

In our protocol, Pb powder was added on top of the gently packed polycrystalline material and immersed in Daphne 7373 oil which served as the PTM. We chose to employ Pb powder instead of wire in an attempt to better represent the local pressure the powdered sample was experiencing. However, we were unable to develop a protocol where we could disperse the Pb powder amongst the packed polycrystalline material while minimizing sample loss in the process. Since the magnetometry measurements are typically normalized per mole of magnetic ions, we decided to prioritize a more accurate measurement of the sample mass over the quantification of pressure since there is more systematic uncertainty in pressure determination. 

This choice has consequences for the appearance of perfect diamagnetism in Pb during our measurements. In a powder, each grain may experience a different pressure, especially at the boundary of the sample holder where pistons are in contact with the sample. This possibly results in a distribution of $T_c$ values and thus likely explains the observed broadening in the superconducting transition in our data [See Fig.~\ref{fig:si-pbtc}]. In passing, we note that this offers another way to inspect the hydrostaticity, this time at the boundary of the pressure cell. However, it complicates the determination $T_c$. It is standard practice to fit a step function to the superconducting transition and assign the midpoint of as $T_c$. The broadening makes this approach less reliable. Since pressure changes from a reference state are the relevant quantity, we chose to use the last (high-temperature) data point before the onset of diamagnetism to determine $T_c$, and subsequently use Eq.~\ref{eq:pressure-corrected} to obtain a lower bound on the actual pressure experienced by the sample. 

\newpage
\noindent{\bf 3. Reproducibility of our susceptibility measurements} \\

\begin{figure}[h]
	    \centering
	    \includegraphics[width = 0.9\linewidth]{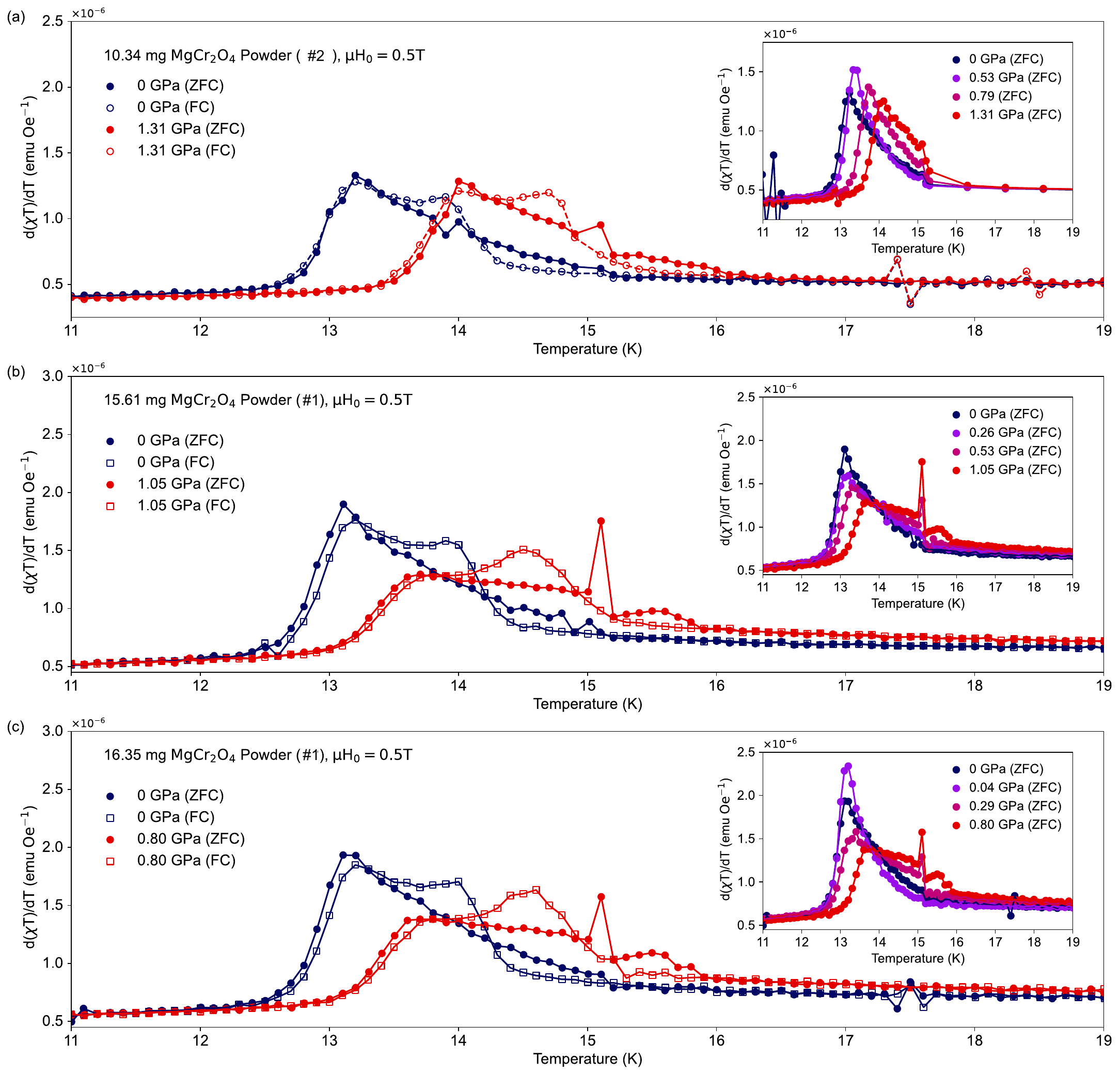}
	    \caption{Replication of our results across three different experiments. Derivative $\partial [\chi\!\cdot\!T]/\partial T\vert_H$ in the vicinity to $T_{\rm N}$ extracted from three distinct $\chi(T)$ measurement run on two different samples. The labels (a--c) match the panels of Fig.~\ref{fig:si-pbtc}, with the experiment presented in the main text [Fig.~\ref{fig:1}] corresponding to panel (c). The insets show four different applied pressures per run resulting from our three-step compression protocol. Overall, the results are consistent across replicates for the same sample in addition to being reproducible across samples with the primary variance originating from the magnitudes of the respective peaks in $\partial [\chi\!\cdot\!T]/\partial T\vert_H$ which is expected to have some sample dependence if associated with variations in domain population. 
	    }
	   \label{fig:si-replicates}
	\end{figure}

 In Fig. \ref{fig:si-replicates}, we compare complete magnetometry measurements performed on sample $\#2$ [Fig. \ref{fig:si-replicates}(a)] with two distinct measurements performed on sample $\#1$ [Fig. \ref{fig:si-replicates}(b--c)], including what is presented in the main text. For the sake of comparison, some intermediate compressions measured with temperature steps of 0.01 K steps between $T=14.8$ and $15.3$~K were rebinned to 0.1K steps to match the later measurements [inset of Figure \ref{fig:si-replicates}(a)]. Although sampling slightly different pressures, replicate experiments ran on sample $\#1$ [Fig. \ref{fig:si-replicates}(b--c)] yielded nearly identical responses, demonstrating the reproducibility of our results. Furthermore, we note that changes in the magnetic peak with pressure are consistent across samples. Subtle variations such as the relative peak intensity may result from domain population, which is expected to be sample dependent. However, when comparing the results shown in Fig.~\ref{fig:si-replicates}(a) with Fig.~\ref{fig:si-replicates}(b-c) it is clear that the hysteresis and the shift in $T_{\rm N}$ with pressure, as well as the appearance of the discontinuity and the broad peak between $T=15$ and $16$~K discussed previously, are consistent across all experiments, regardless of the sample batch employed.

\newpage
\noindent{\bf 4. Discontinuity in $\partial [\chi\!\cdot\!T]/\partial T\vert_H$ under pressure} \\

\begin{figure}[h]
	    \centering
	    \includegraphics[width = 0.9\linewidth]{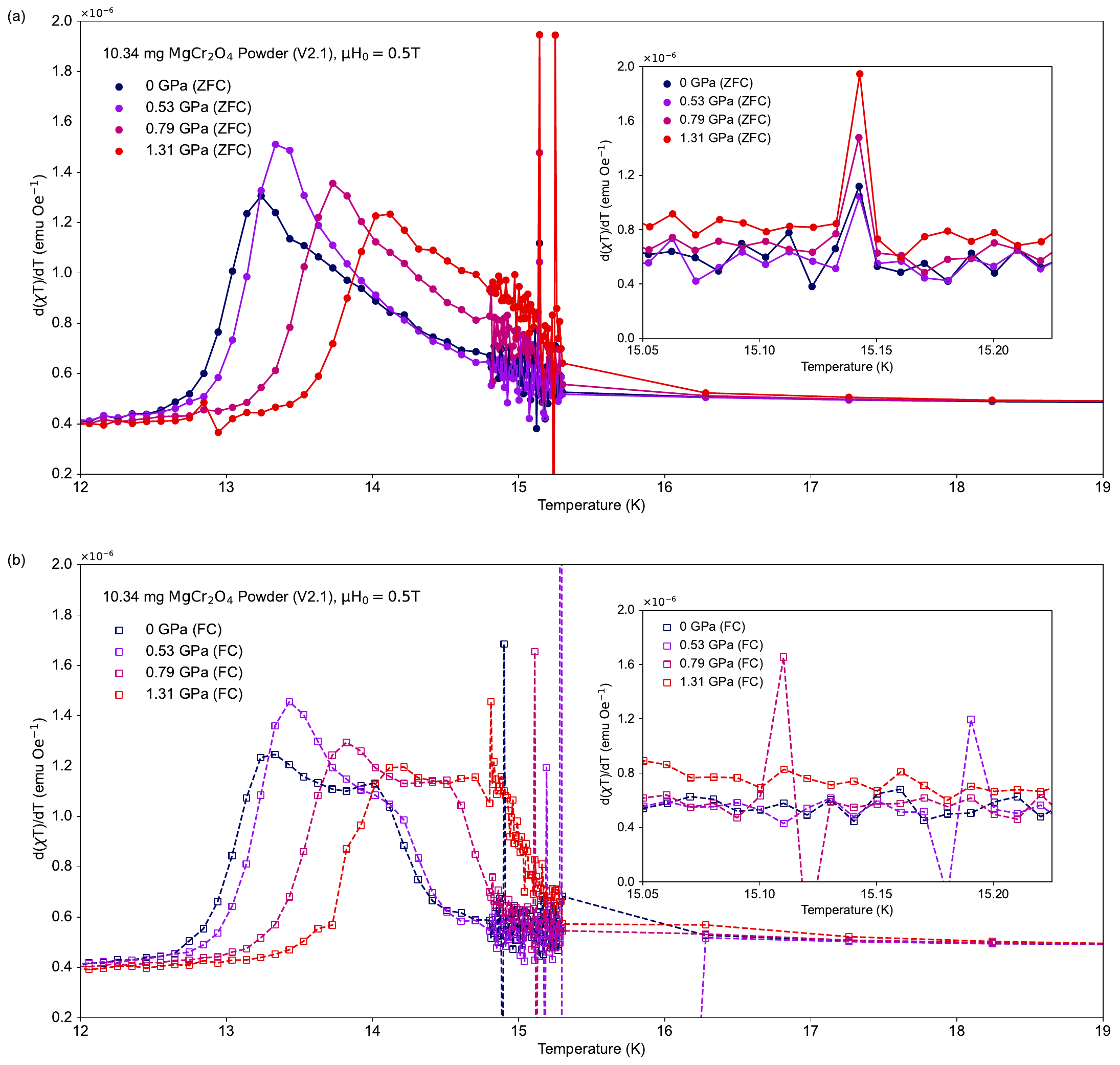}
	    \caption{High temperature resolution scans of $\partial [\chi\!\cdot\!T]/\partial T\vert_H$ in the vicinity to $T_{\rm N}$ in (a) zero-field cooled (warming) and (b) field-cooled conditions for sample $\#2$.  The insets shows the region between $T=15.0$~K and $T=15.2$~K where the sharp discontinuity is observed.}
	    \label{fig:si-discontinuity}
	\end{figure}

In addition to the expected peak indicating the transition at around $T=13$~K, a second very sharp peak at around $T=15$~K was observed to emerge with increasing applications of pressure in the $\partial [\chi\!\cdot\!T]/\partial T\vert_H$  data. This feature was initially assumed to be an instrument error or noise, however, given the reproducibility of the feature, the region of the 'peak' between  $T=14.8$~ K and $T=15.3$~K, was scanned with very fine temperature steps of 0.01K steps to capture it with more data points [See Fig.~\ref{fig:si-discontinuity}]. This feature changes in amplitude with pressure and demonstrates a field dependence, as it does not appear in field-cooled measurements. However, even with this resolution, this `peak' is captured by just a single point in the derivative.

\newpage
\noindent{\bf 5. Field and thermal dependence of the specific heat around $T_{\rm N}$} \\

	\begin{figure}[h]
	    \centering
	    \includegraphics[width = 0.7\linewidth]{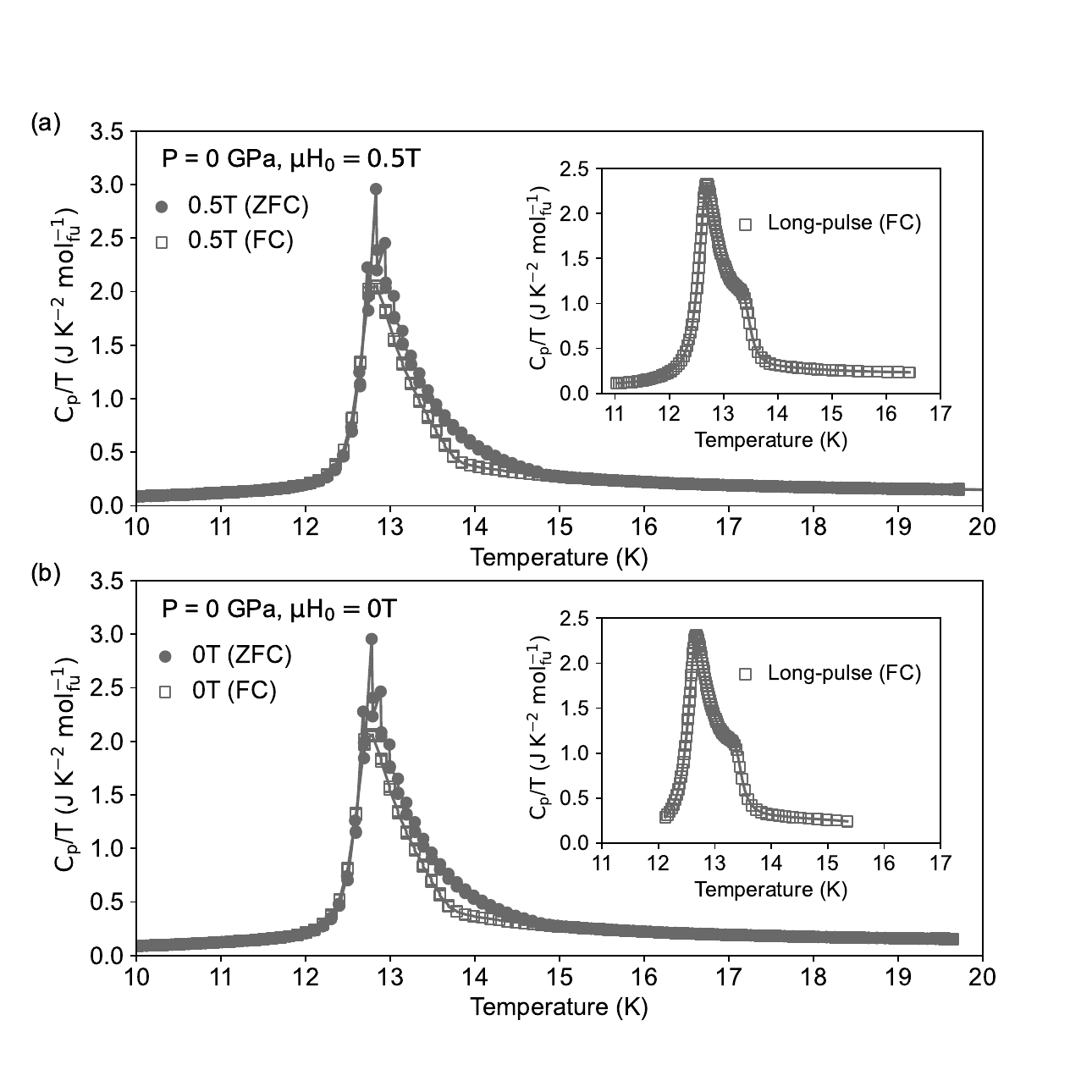}
	    \caption{Comparison of heat-capacity measurements performed with and without an applied field of $\mu_0H=0.5$~T (to match with magnetometry measurements) showing essentially indistinguishable results and confirming the thermal origin (rather than the magnetic field origin) of the hysteresis phenomena observed in these measurements.}
	    \label{fig:si-hc}
	\end{figure}

\newpage
\noindent{\bf 6. Structural Bragg Peaks}\\

	\begin{figure}[h!]
	    \centering
	    \includegraphics[width = 0.5\linewidth]{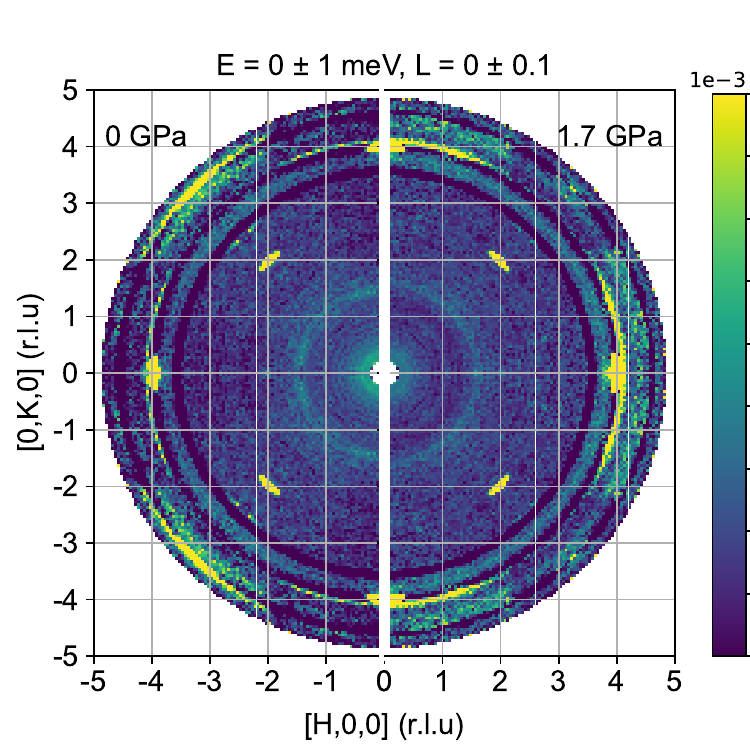}
	    \caption{Slices through the symmetrized elastic neutron scattering data collected above the Néel ordering temperature at $T = 20$~K, comparing ambient pressure with results for $P = 1.7$~GPa.}
	    \label{fig:si-structure}
	\end{figure}

\end{document}